\documentclass{aa}  

\usepackage{graphicx}
\usepackage{txfonts,textcomp}
\usepackage[allcolors=blue]{hyperref}
\newcommand{\dif}[1]{\mathrm{d}#1\,} 
\newcommand{\pderiv}[2]{\frac{\partial #1}{\partial #2}} 

\newcommand{\vr}{\vec{r}}
\newcommand{\ve}{\vec{e}}
\newcommand{\vu}{\vec{u}}
\newcommand{\vn}{\vec{n}}
\newcommand{\vpi}{\vec{\pi}}
\newcommand{\vom}{\vec{\omega}}

\newcommand{\cchi}{\cos\chi}
\newcommand{\schi}{\sin\chi}
\newcommand{\cphi}{\cos\phi}
\newcommand{\sphi}{\sin\phi}
\newcommand{\cmu}{\cos\mu}
\newcommand{\smu}{\sin\mu}
\newcommand{\tmu}{\tan\mu}
\newcommand{\crho}{\cos\rho}

\newcommand{\schisq}{\sin^2\chi}

\newcommand{\cmusq}{\cos^2\mu}

\newcommand{\bphi}{\bar\phi}

\newcommand{\vB}{\vec{B}}
\newcommand{\vb}{\vec{b}}
\newcommand{\vx}{\vec{x}}

\begin{document}

   \title{Geometrical envelopes of fast radio bursts}


\author{G. Voisin\inst{1}\fnmsep\thanks{Secondary email: astro.guillaume.voisin@gmail.com}
}
 
\institute{%
        LUTh, Observatoire de Paris, PSL Research University, CNRS, Université de Paris, Sorbonne Université, 5 place Jules Janssen, 92190 Meudon, France \\
        \email{guillaume.voisin@obspm.fr}
        }

   \date{Received September 15, 1996; accepted March 16, 1997}

 
  \abstract
   {}
   {Assuming fast radio bursts (FRBs) are produced by matter travelling ultra-relativistically in a localised region of a smooth bundle of streamlines, we study the constraints applied by geometry to the morphology and polarisation of the burst in time and frequency independently of the intrinsic radiative process. }
   {We express the problem only in terms of the local properties of direction and curvature of a streamline. This allowed us to cast the general results to any desired geometry. By applying this framework to two geometries inspired by pulsar and magnetar magnetospheres, we namely illustrate the dipolar polar-cap region and a magnetic dipole with an additional toroidal component.}
   {Geometry constrains bursts to occur within an envelope in the frequency versus time plane (dynamic spectrum). This envelope notably characterises spectral occupancy and frequency drifts (both burst-to-burst and within an individual burst). We illustrate how one can simulate bursts by specifying some basic properties of an intrinsic emission process. In particular we show that the typical properties of one-off bursts can be produced in polar-cap geometry by a star with a spin period $>1$s, while bursts from repeating sources are better accounted for with an additional strong toroidal component and a sub-second spin period. }
   {We propose that a relationship between burst morphologies and the properties of the source, such as its spin period and magnetospheric properties, can be established at least qualitatively based on geometrical considerations. Our results favour models where repeaters are younger and faster magnetars with highly twisted magnetospheres. }


   \maketitle
%

\section{Introduction \label{sec:intro}}

As observations of fast radio bursts (FRBs) greatly increased in the last decade \citep[e.g.][]{petroff_fast_2022}, their properties became increasingly diverse while patterns emerged. 
First, their occurrence statistics is now divided between repeaters and apparent non-repeaters, which are sources that have produced only one event so far. Repeaters themselves display highly variable rates and regularity, with some alternating between burst `storms' and quiet periods \citep{lanman_sudden_2022}, and they sometimes display periodic activity windows such as the well-studied FRB 180916  \citep{collaboration_periodic_2020} and FRB 121102 \citep{rajwade_possible_2020}. The latter has been shown to be Poisson-distributed within windows once events with millisecond separation are excluded \citep{cruces_repeating_2021}.
Second, their detailed morphologies in the time-frequency plane (the dynamic spectrum) vary from one burst to another, including between bursts from the same source \citep{pleunis_fast_2021}. However, some general characteristics can be distinguished from the available samples. Repeaters produce spectrally narrower bursts of a longer duration than one-off events which tend to be short wide-band events \citep{pleunis_fast_2021, chamma_broad_2022}. Importantly, for our discussion, repeaters often produce short sequences of sub-bursts and the central frequency of which drifts downwards. The fact that sub-bursts are drifting downwards has been dubbed the `sad-trombone effect' \citep{hessels_frb_2019}.
A very short substructure has been reported at a microsecond scale \citep{farah_frb_2018, nimmo_highly_2021, majid_bright_2021}, and there is at least one occurrence of quasi-periodic sub-bursts \citep{andersen_sub-second_2022}. 
Third, polarisation is usually linear; although, occasional circular polarisation has been observed in some sources (see sec. 5.9 of \citep{caleb_decade_2021} for a review). Variability in the polarisation angle (PA) \citep[e.g.][]{hilmarsson_polarization_2021,nimmo_highly_2021} as well as  in the rotation measure (RM) and dispersion measure (DM) have been observed burst to burst \cite[e.g.][]{michilli_extreme_2018, anna-thomas_highly_2022, mckinven_large_2022}. For repeaters in particular, a flat PA profile is observed within bursts with variations in values from one burst to another. A notable exception to the flat profile is FRB180301 \citep{luo_diverse_2020}, where PA swings within bursts that are reminiscing of pulsar profiles have been observed. Variations in RM and DM have been proposed to result from propagations through a turbulent magneto-active environment \citep{michilli_extreme_2018} surrounding the source, while polarisation swings are thought to be intrinsic to the source \citep{luo_diverse_2020}. 

Many models have been proposed to explain FRBs. As we are mostly interested in geometry in this paper, it is fairly independent from the detailed emission and plasma mechanisms involved. However, it was written with the idea that FRBs are produced in the environment of a neutron star (NS). These models can be divided into two categories: magnetospheric models \citep[e.g.][]{popov_millisecond_2013, cordes_supergiant_2016, kumar_fast_2017, wadiasingh_repeating_2019, wadiasingh_fast_2020, lu_unified_2020, lyubarsky_fast_2020}, where emission takes place within a few light-cylinder radii of the NS, and wind models where the distance is much larger. Most wind models consider the propagation of a shock wave producing coherent cyclotron maser emission \citep[e.g.][]{popov_millisecond_2013, lyubarsky_model_2014, metzger_millisecond_2017, beloborodov_flaring_2017, beloborodov_blast_2020, yuan_plasmoid_2020}.
An alternative category of models involves the interaction of small bodies with the magnetosphere or wind of a NS. In the former case, the energy source is gravitational as the asteroid falls within the magnetosphere \citep[e.g.][]{dai_repeating_2016, bagchi_unified_2017, smallwood_investigation_2019, dai_periodic_2020, dai_magnetar-asteroid_2020}; whereas, in the latter case, energy is taken from the wind as the object orbits the star \citep{mottez_radio_2014, mottez_repeating_2020,voisin_periodic_2021}. 

In most cases, the NS is a magnetar and energy is released in star quakes occurring under magnetic stress following the theory of magnetar outbursts \citep{thompson_soft_1996, perna_unified_2011} (with the exception of falling asteroids). Models mainly differ  by the mechanism by which power is converted into radio waves and its location. 
Importantly, in every case the trigger of FRBs is intrinsically unpredictable, and the process involves an ultra-relativistic plasma. 

This work relies on the idea that due to their very short duration, FRB emissions must be very localised in space. In addition, their high flux favours models with highly collimated emission in order to keep a sufficiently low energy budget. This property is included in many if not all of the currently proposed models at various degrees. Thus, in Sec. \ref{sec:model} we develop a general model based on the assumption that FRBs result from the emission of a portion of streamline of ultra-relativistically moving plasma which is localised enough for the geometry of the visible segments of streamlines to be described by their tangent and curvature only. Further, assuming that the properties of streamlines vary slowly compared to the size of the emission region, we reduced the local bundle to a single streamline endowed with an effective emission angle that is a free parameter encoding both the intrinsic angle of the emission process and the effect of divergence of field lines within the bundle. As a general feature, we consider that streamlines are fixed in the rotating frame of the star and that the co-rotating velocity can be neglected in front of the speed of light. In order to map geometry into the frequency domain, we make the assumption of radius-to-frequency mapping (\citet[e.g.][]{cordes_observational_1978, lyutikov_radius--frequency_2020} and Sec. \ref{sec:rfmap}), whereby emission frequency decreases according to a power law of altitude. Assuming a relation between polarisation and geometry, we can also derive the PA envelope. For this work, we have adapted the rotating-vector model \citep{radhakrishnan_magnetic_1969, komesaroff_possible_1970, petri_polarized_2017} which is commonly used to model pulsar PA profiles and where radiation is assumed to be linearly polarised within the plane of curvature. 

The result of this simple description is to provide a visibility envelope in the time-frequency plane together with the PA envelope for any set of local geometrical parameters. Within the envelopes, FRB properties are unspecified. In particular, the exact spectrum of FRBs within the envelope depends on the exact plasma distribution and intrinsic emission mechanism, while the time distribution depends on an external trigger (for example a star quake or an asteroid) independent from local geometry. Further, we show in Sec. \ref{sec:profiles} how to model synthetic bursts using intrinsic emission profiles, which must be derived from a particular process. 

In order to connect bursts together occurring at different epochs in different regions of the source, one needs to adopt a global description of the geometry from which local properties can be derived. In section \ref{sec:applications}, we explain how we applied our model to the geometry of a dipolar polar cap, defined here as the bundle of field lines that reach the light cylinder radius (the radius at which co-rotation velocity equals the speed of light) of a magnetic dipole. We then investigate the effects of a toroidal component added to this dipole. The first geometry is common in the study of pulsar magnetospheres at low altitudes compared to the light cylinder radius, while the second is motivated by the expectation of a toroidal component in magnetar magnetospheres \citep[e.g.][]{thompson_electrodynamics_2002, perna_unified_2011, barrere_new_2022} and by the fact that it empirically allowed us to reproduce some observational properties of FRBs such as downward drifting sub-bursts. 
We give examples of synthetic bursts for both geometries in Figs. \ref{fig:burstJ1935} and  \ref{fig:bursttoro}. 

We note that, in principle, our approach could apply to other categories of the model mentioned above, such as shock waves or orbiting asteroids in a NS wind, provided the appropriate geometry. The examples chosen here also present the advantage of allowing a simple yet insightful analytic treatment. 

The paper is organised as follows: Sec. \ref{sec:model} develops the models for time-frequency and polarisation-angle envelopes, Sec. \ref{sec:bursts} extends the model to bursts, Sec. \ref{sec:applications} explores the properties of the model when applied to the two geometries described above, and we discuss in Sec. \ref{sec:discussion} how this gives us insight into the relationship between burst morphology and the repeating or apparently non-repeating nature of the source. We summarise in Sec. \ref{sec:conclusion}.

\section{Model of geometrical envelopes \label{sec:model}}

\subsection{Bundle reduction \label{sec:bundlereduc}}
\begin{figure}
        \centering
        \includegraphics[width=0.45\textwidth]{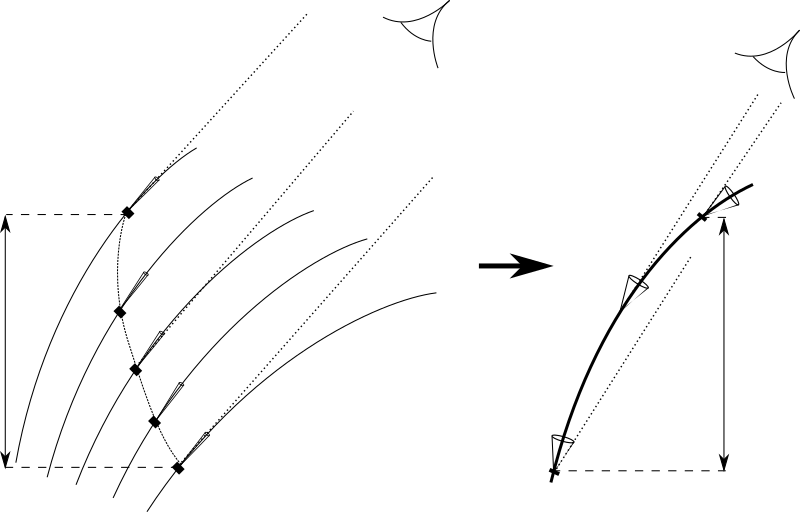}
        \caption{Sketch representing the reduction of a bundle of streamlines to a single line with an effective emission angle. Left: Bundle of lines with narrow emission angles (cones) that all have a tiny portion (thickness of the tick at the base of the cone) simultaneously visible by the observer (light goes along dotted lines). Right: This particular bundle was reduced to a single line with a wider emission angle such that the length of line along the vertical (altitude) axis visible by the observer is the same as the cumulated portions of lines of the bundle on the left.  }
        \label{fig:sketchreduc}
\end{figure}

We assume that emission is very collimated, which is a natural consequence of utlra-relativistic motion independently of the physics of the emission process. We model the beam by a cone characterised by its axis given by the local velocity (tangent to the streamline) and opening angle. The segment of streamline visible by an observer at a given instant is the set of all points the cone of which contains the observer. 
For simplicity we consider that all streamlines of the local emission region share the same opening angle. The bundle of visible streamlines is made of all lines which have a visible segment during a burst or a close sequence of bursts. The bundle is assumed to be small enough so that all lines share approximately the same curvature and may only be slightly shifted or rotated with respect to one another (see Fig. \ref{fig:sketchreduc}). As a result, each visible segment is at a different altitude such that their union covers a wider altitude range than each of them taken separately. This is important due to the assumption of radius-to-frequency mapping. 

\begin{figure}
        \centering
        \includegraphics[width=0.45\textwidth]{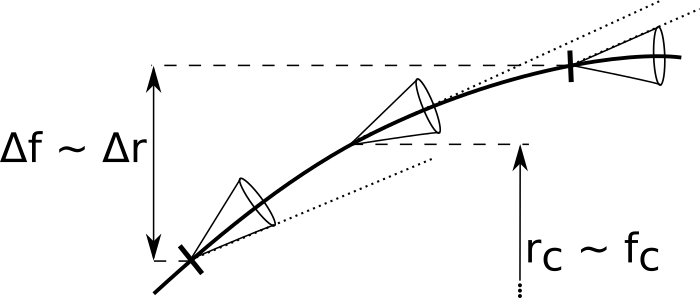}
        \caption{Illustration of radius-to-frequency mapping on a streamline (thick line) endowed with an effective emission cone. Ticks are located where light rays (dotted lines) graze the local cone, thus delimiting the visible segment. By virtue of radius-to-frequency mapping, the projection of the segment along the radial (vertical) axis defines the frequency bandwidth, while the centre of the segment gives the central frequency. }
        \label{fig:sketchr2freq}
\end{figure}

Since we ignore the contours of the visible bundle as well as the intrinsic emission angle, considering in addition that these quantities may be burst-dependent, we propose to reduce the problem to a single line and absorb the unknowns in an effective emission angle, Figs. \ref{fig:sketchreduc} and \ref{fig:sketchr2freq}. This angle is a priori burst-dependent, meaning that each event may have a different value as it may come from a different bundle or have different plasma properties. 

In this work we consider that close sub-burst sequences, typically with components separated by less than a millisecond, constitute a single event with regard to the geometrical quantities involved. In particular, all components share the same effective opening angle. The millisecond threshold is justified by the fact that events more widely separated appear to be uncorrelated in FRB121102 \citep{cruces_repeating_2021}.

In general, a sufficiently smooth and narrow bundle can be approximated at leading order by a single line endowed with an opening angle characterising the emission process. Since we are primarily interested into dynamic spectra, departures from this approximation may lead either to an incorrect bandwidth or duration associated with the visible part of the bundle. Below we explain why our approach is much less sensitive to duration than bandwidth. This means that even if a bundle departs significantly from the single-line limit, it produces observational changes primarily along a single dimension, bandwidth, which can in principle be absorbed by fitting a single effective parameter, the effective opening angle. As a consequence we expect this model to be relatively robust to departures from the single-line approximation.

There is no unique way to define an effective opening angle when the bundle departs from a single line. In this paragraph we discuss two cases which illustrate the different constraints.
 The effective angle may be chosen such that the visible segment covers the same altitude range as the union of individual segments (see Fig. \ref{fig:sketchreduc}). Considering radius-to-frequency mapping, this choice allows us to reproduce the correct bandwidth, Fig. \ref{fig:sketchr2freq}. 
An alternative choice would be to choose the angle in order to reproduce a certain duration. Indeed, the wider the angle the longer the line remains visible as the source rotates. 
However we consider in this work that the duration of bursts is trigger-limited in the sense that it is significantly shorter than the duration of the visibility of the emitting region, and therefore is not limited by it. If that were the case a bundle becoming invisible could in principle be replaced by a neighbouring bundle as the star rotates, thus breaking the locality hypothesis in the sense that a burst could not be explained using a single bundle. Thus, as long as bursts are trigger-limited the duration of the bundle's visibility is not a very constraining quantity.
On the other hand, radius-to-frequency mapping almost imposes that the entire bandwidth allowed by the visible region at a given time within a burst be filled unless the radiating plasma is distributed irregularly on scales smaller than the size of the visible region. Thus, in case of departure from the single-line limit the effective opening angle is primarily adjusted to fit the correct bandwidth rather than duration. 

\subsection{Streamline local description}

Locally a streamline can be described in the Serret-Frenet orthonormal frame $F_0 =(\ve_1|_0, \ve_2|_0, \ve_3|_0)$ at $\vr$ by 
\begin{equation}
\label{eq:OM}
\vec{OM} = \kappa^{-1}\left[\zeta (1 - \frac{\zeta^2}{6}) \ve_1|_0 + \frac{\zeta^2}{2}\left(1 + \frac{\kappa'}{\kappa}\frac{\zeta}{3}\right) \ve_2|_0 + \bar\tau \frac{\zeta^3}{6} \ve_3|_0 \right] + \bigcirc\left(\zeta^4\right),
\end{equation}
where $\zeta = \kappa s$ is the curvilinear abscissa normalised by curvature radius $R_c = 1/\kappa$, $s$ is the canonical curvilinear abscissa such that $\dif{\vec{OM}}/\dif{s}(s=0) =\ve_1|_0$, $\kappa' = \dif{\kappa}/\dif{\zeta}(\zeta=0)$ and $\bar\tau = \tau\kappa$ is the torsion $\tau$ normalised by the curvature radius. In the following, we study the simplest case beyond the straight line: we assume no torsion and uniform curvature, that is $\kappa' = \tau = 0$. It follows that the tangent vector along the line is 
\begin{equation}
\label{eq:e1}
\ve_1(\zeta) =  (1 - \frac{\zeta^2}{2}) \ve_1|_0 + \zeta\ve_2|_0.
\end{equation}

\subsection{Frames}
The source and the observer are considered fixed with respect to one another. They share a common frame $F = (\ve_x, \ve_y, \ve_z)$ where $\ve_z$ is aligned with the spin axis of the source. 

The emitting bundle is fixed with respect to a rotating frame $F'=(\ve_x', \ve_y', \ve_z' = \ve_z)$ and, unless otherwise specified, the position of the bundle $\vr$ is fixed with respect to $F'$. The rotating frame $F'$ is the image of $F$ by a rotation of axis $\ve_z$ and angle $\phi = \phi_0 + \omega t$ where $\omega = 2\pi/P$ is the angular velocity of the source, that is
\begin{eqnarray}
        \ve_x' & = & \cphi \ve_x + \sphi \ve_y, \\
        \ve_y' & = & -\sphi \ve_x + \cphi \ve_y, \\
        \ve_z' & = & \ve_z. 
\end{eqnarray}
Angle $\phi_0$ is such that the tangent vector to the streamline at $\vr$ is the image of $\ve_z$ by a rotation of angle $\chi \in [0,\pi]$ around axis $\ve_y'$,
\begin{equation}
\ve_1|_0 =  \schi \ve_x' + \cchi\ve_z.
\end{equation}
The vector $\ve_1|_0$ can be completed to form a frame $F''= (\ve_x'', \ve_y'', \ve_z'' = \ve_1|_0)$ image of $F'$. The Serret-Frenet frame $F_0$ associated to the streamline at $\vr$ is then the image of $F''$ by a rotation of angle $\mu$ around axis $\ve_1|_0=\ve_z''$.

\begin{figure}
        \centering
        \includegraphics[width=0.3\textwidth]{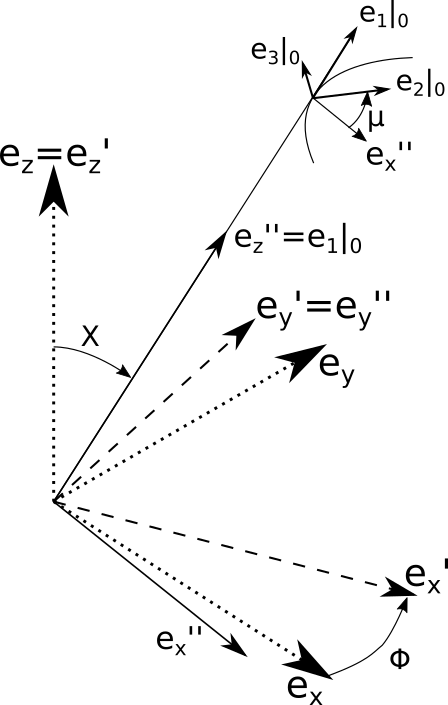}
        \caption{Illustration of the relations between the frames $F$, $F'$, $F''$, and $F_0$, Eqs. \eqref{eq:frameF}-\eqref{eq:frameF0}. The curved line represents a small portion of streamline curved in the plane $(\ve_1|_0, \ve_2|_0)$. }
        \label{fig:frames}
\end{figure}

We summarise the transformations in Fig. \ref{fig:frames} and below,
\begin{eqnarray}
\label{eq:frameF}
        F & = &  (\ve_x, \ve_y, \ve_z), \\
        F' & = & R_{\ve_z, \phi}(F) =  (\ve_x', \ve_y', \ve_z' = \ve_z),\\
        F'' & = & R_{\ve_y', \chi}(F') =  (\ve_x'', \ve_y'' = \ve_y', \ve_z'' = \ve_1|_0),\\
        F_0 & = & R_{\ve_z'', \mu}(F'') = (\ve_1|_0, \ve_2|_0, \ve_3|_0).
        \label{eq:frameF0}
\end{eqnarray}

It follows that the Serret-Frenet frame $F_0$ is related to the observer's frame $F$ by
\begin{eqnarray}
\label{eq:e10}
        \ve_1|_0 & = & \schi \cphi \ve_x + \schi\sphi \ve_y + \cchi\ve_z \\
        \ve_2|_0 & = & (\cmu \cchi \cphi - \smu\sphi) \ve_x, \\
        & & + (\cmu \cchi \sphi + \smu\cphi)\ve_y - \cmu\schi \ve_z, \nonumber\\
\label{eq:e30}
        \ve_3|_0 & = & -(\smu \cchi \cphi + \cmu\sphi) \ve_x \\
        & & + (-\smu \cchi \sphi + \cmu\cphi)\ve_y + \smu\schi \ve_z.  \nonumber
\end{eqnarray}
From these equations we see that the inclination $\chi$ and curvature-plane angle $\mu$ can be deduced from the projection of $F_0$ on the spin axis $\ve_z$ by
\begin{eqnarray}
\label{eq:chi}
        \chi & = & \arccos\left(\ve_1|_0\cdot\ve_z\right), \\
        \label{eq:cosmu}
        \cos\mu & = & -\frac{\ve_2|_0\cdot \ve_z}{\schi}, \\
        \label{eq:sinmu}
        \sin\mu & = & \frac{\ve_3|_0\cdot \ve_z}{\schi}.
\end{eqnarray}

\subsection{Visibility region along a streamline in a rotating frame}
Emission in a cone of opening angle $\Omega$ around the local tangent is visible by an observer in a direction given by the unit vector $\vu$ provided that $\gamma < \Omega$ where $\gamma$ is the angle of the tangent with respect to the line of sight. That is, 
\begin{equation}
        \label{eq:boundarysl}
        \cos\gamma \equiv \vu\cdot \ve_1 > \cos\Omega.
\end{equation}
Without a loss in generality, we may choose an observer in the plane $(\ve_x, \ve_z)$ such that 
\begin{equation}
        \label{eq:u}
        \vu = \schi \ve_x + \cchi\ve_z.
\end{equation}
Inserting Eq. \eqref{eq:e1} into Eq. \eqref{eq:boundarysl} and using Eqs. \eqref{eq:e10}-\eqref{eq:e30}, we obtain 
\begin{equation}
\label{eq:boundarysl2}
        u_1 + \zeta u_2 - \frac{\zeta^2}{2}u_1 > \cos\Omega,
\end{equation}
where $(u_1,u_2, u_3)$ are the components $\vu$ relative to $F_0$,
\begin{eqnarray}
\label{eq:u1}
        u_1 & = & \schi^2\cphi + \cchi^2, \\
\label{eq:u2}
        u_2 & = & \schi \left(\cmu\cchi\cphi - \smu\sphi\right) \\
        & & - \cmu\cchi\schi. \nonumber
\end{eqnarray}
In order to have $\ve_1|_0 = \vu$ at $t=0$ we see that we need to choose $\phi_0 = 0$. 

With that choice, the left-hand side of Eq. \eqref{eq:boundarysl2} has a maximum at $\zeta=t=0$ (that is $\vu = \ve_1|_0$), and Eq. \eqref{eq:boundarysl2} describes a segment bounded by the two solutions of $\vu\cdot \ve_1 -\cos\Omega =0$ which is a second-order polynomial in $\zeta$, 
\begin{eqnarray}
\label{eq:zetapm}
        \zeta_{\pm} &=& \zeta_c \pm \frac{\Delta\zeta}{2}, \\
        \zeta_c  & =& -\frac{u_2}{u_1}, \\
        \Delta \zeta & = & 2\frac{\sqrt{\Delta}}{u_1}, \\
\label{eq:Delta}
        \Delta & =& u_2^2 +2u_1(u_1 - \cos\Omega),
\end{eqnarray}
where $\zeta_c$ is the coordinate of the centre of the segment, $\Delta\zeta$ its length, and $\Delta$ is the determinant of the polynomial. They are all functions of time via its dependence on $\phi$. At time $\phi=\omega t$, a segment is visible only if a solution exist, that is if $\Delta > 0$. Therefore solutions to $\Delta > 0$ provide us with the duration during which a given bundle is visible. Here we assume that $\phi \ll 1$, that is the bundle is visible for much less that the spin period, which allows to Taylor expand neglecting terms of order $\bigcirc(\phi^3)$. Similarly, we neglect terms of order $\bigcirc(\Omega^3)$. The bundle is then visible provided that
\begin{equation}
        1 - \left(1+\frac{\cmusq}{\delta_\Omega}\right)\schisq \frac{\phi^2}{2} > 0,
\end{equation}
where $\delta_\Omega = 1 - \cos\Omega$. This equation is fulfilled provided that $\phi \in ]-\Delta\phi/2, +\Delta\phi/2[$ with 
\begin{equation}
\label{eq:dphi}
        \Delta\phi = \frac{2}{\schi} \sqrt{\frac{2\delta_\Omega}{\delta_\Omega + \cmusq}} = 2\frac{\Omega}{\schi|\cmu|} + \bigcirc(\Omega^3),
\end{equation}
where we Taylor-expanded to leading order in $\Omega$ on the right hand-side, showing that the assumption $\Delta\phi \ll 1$ is consistent if $\Omega \ll 1$. However, this assumption breaks down if the spin axis faces the observer, that is $\sin\chi = 0$, or if the streamline is curved in the plane parallel to the spin equator, that is $\cmu = 0$. The bundle remains visible for a time $\Delta t = \Delta\phi/\omega$, 
\begin{equation}
\label{eq:dt}
        \Delta t = \frac{2 P}{|\cmu|\schi} \frac{\Omega}{2\pi} + \bigcirc(\Omega^3),
\end{equation} 
where $P = 2\pi/\omega$ is the spin period. 

To leading order in $\Omega$, Eq. \eqref{eq:zetapm} can conveniently be expressed as a function of the normalised spin phase $\bphi \equiv \phi/(\Delta\phi/2)$,
\begin{equation}
        \label{eq:zetapm2}
        \zeta_\pm(\bphi) = \Omega\left[-\bphi \epsilon \tan\mu \pm \sqrt{1-\bphi^2}\right],
\end{equation}
where $\bphi \in [-1,1]$, and $\epsilon = \mathrm{sign}(\cos\mu)$.

\subsection{Burst envelopes from radius-to-frequency mapping \label{sec:rfmap}}
We assume a relation between altitude $r$ and the emission frequency $f$ of the form \citep[e.g.][]{cordes_observational_1978, lyutikov_radius--frequency_2020}
\begin{equation}
\label{eq:fr}
        f(r) = F \|\vr\|^{-\beta},
\end{equation}
where $\beta$ is an unknown index, and $F$ is a normalisation constant. 

The curvilinear abscissa $\zeta$ maps to the radius $r(\zeta)$ according to the following projection, 
\begin{equation}
\label{eq:r}
        r(\zeta) = r + \vec{OM}(\zeta)\cdot \ve_r' 
\end{equation}
where $r = \|\vr\|$ and $\ve_r' = \vr/r$, remembering that $\vr$ is defined relatively to $F'$.

Keeping only leading order terms in $\zeta \sim \Omega$ (see Eq. \eqref{eq:zetapm2}) we obtain 
\begin{equation}
\label{eq:r2}
        r(\zeta) = r + \zeta \kappa^{-1}\crho,
\end{equation} 
where $\crho \equiv \ve_r' \cdot \ve_1|_0$.

Radius-to-frequency mapping allows us to map a bounded region in radius to a bounded region in the frequency domain. If times of arrival $t_a$ measured by the observer can be associated to these boundaries, it defines an observable envelope within which bursts are visible. 
We define the burst envelope at $\vr$ as the region $\{t_a \in [t_a{}_-,t_a{}_+]; f\in[f_-(t_a), f_+(t_a)]\}_{\vr}$ where 
\begin{eqnarray}
\label{eq:fpm}
f_\pm(t_a) & = & f\left(\max\left(r_\mp(t_a), R_*\right)\right), 
\end{eqnarray}
where $R_*$ is the radius of the star, $r_\pm$ are the lower and upper radii of the visible region, and $f()$ is defined in Eq. \eqref{eq:fr}. We use the fiducial value $R_* = 12$ km in the figures of this paper. This definition incorporates the fact that the emission region cannot extend below the surface of the star, which translates into an upper frequency cut-off (for $\beta>0$).

\subsection{Propagation}\label{sec:propagation}
\newcommand{\bkappa}{\bar\kappa}

We shall consider the case where the streamline is a bright filament in a rotating frame, that is to say it is continuously emitting. Due to the curvature and beaming angle, an observer sees the segment delimited by Eq. \eqref{eq:zetapm2}, and the maximum duration is limited to Eq. \eqref{eq:dt} by the rotation of the frame. We call the envelope thus defined instantaneous visibility envelope, and provide a detailed treatment in appendix \ref{ap:instenv}.
In practice we are interested in cases where emission is short compared to the duration of the envelope. In this case the instantaneous envelope is the limit of negligible propagation time through the segment, that is the visible segment is lighten everywhere at once. 

Now, we shall consider an emitting element propagating along a streamline. What we call emitting element is to be seen as a signal that switches on emission from the location it is crossing. In practice, it may be a moving plasma or a wave front. For definiteness we assume it propagates from lower to higher altitudes. The emitting element emits towards the observer only when it crosses the visible segment defined in Eq. \eqref{eq:zetapm2} that is itself moving and changing length with time. For example, the visible segment may very well move up faster than the speed of light, in which case emission is seen only from those elements that were injected on the line before the segment became visible at all (that is when $|\bar\phi| > 1$) and that are being caught up by the it. A point $(t_a,f)$ in the time-frequency plane is within the envelope if there exists an injection time $t_i$ such that the emitting element belongs to the visible segment when it reaches this point. The remainder of this section details and develops this definition. We recall that the co-rotation velocity is neglected compared to the speed of light.

Assuming an emitting element travels along a streamline at a velocity $v$, its abscissa can be locally defined as 
\begin{equation}
\label{eq:propagzeta}
\zeta_p(t,t_i) = \kappa s =  \bkappa \frac{v}{c} \omega(t - t_i),
\end{equation}
where $t_i$ is the injection time defined as the instant when the element crosses the centre of the emission region at $\zeta(t_i) = 0$, and we have introduced the reduced curvature $\bkappa = \kappa R_L$ where $R_L =c/\omega$ is the light-cylinder radius of a rotating star.

In this section, we use the notation 
\begin{eqnarray}
\label{eq:propagf}
        f(t,t_i) &\equiv& f\left(\max\left(r\left(\zeta_p\left(t,t_i\right)\right), R_*\right)\right) \\
        &  = & \min\left(\frac{F}{\left(r + \crho v(t-t_i)\right)^{\beta}}, \frac{F}{R_*^{\beta}}\right), \nonumber
\end{eqnarray}
which describes the emission frequency at $t$ of an element injected at $t_i$. It derives from the combination of Eqs. \eqref{eq:fr},\eqref{eq:r2}, \eqref{eq:propagzeta} with the restriction to altitude above the surface of the star as in Eq. \eqref{eq:fpm}.

In order to see the importance of propagation, we can estimate the propagation time through the segment assuming negligible rotation. In that case the emission front travels through the entire $\Delta \zeta = 2\Omega$ of Eq. \eqref{eq:zetapm2}. Using Eq. \eqref{eq:propagzeta}, we get $\Delta t_p = 2\Omega c/\bkappa v \omega \sim \Omega P/\bkappa\pi \sim \Delta t$, implying that the condition $\Delta t_p \ll \Delta t$ characterising the domain of the validity of instantaneous envelopes (see above) can be violated and that propagation has to be considered in determining the visibility. In other words, the visible segment can move along the line at a similar speed as the trigger. 

\subsubsection{Burst envelope accounting for finite propagation velocity}
Here we consider that $v=c$ in Eq. \eqref{eq:propagzeta}. The condition of visibility of an emitting element is given by
\begin{equation}
\label{eq:propagvis}
        \zeta_-(t) \leq \zeta_p(t,t_i) \leq \zeta_+(t),
\end{equation} 
where $\zeta_\pm(t)$ is defined in Eq. \eqref{eq:zetapm2}. The solution of these inequalities is of the form $t \in [t_-(t_i), t_+(t_i)]$ and is detailed in appendix \ref{ap:propagvis} where it is shown that
\begin{equation}
\label{eq:propagphipm}
        t_\pm(t_i) = \frac{\Delta t}{2}\frac{a bt_i \pm \sqrt{1 + a^2 - b^2t_i^2}}{1+a^2} \equiv t_c(t_i) \pm \frac{\Delta t_p(t_i)}{2},
\end{equation}
where $a = \epsilon(\bkappa /(\sin\chi\cos\mu) + \tan\mu)$ and $b = \omega  \bkappa /\Omega$, and $\Delta t$ is from Eq. \eqref{eq:dt}. Moreover, we introduce the centre $t_c(t_i)$ and the duration $\Delta t_p(t_i)$ of the visible time interval. 
For small or moderate value of $a$ and $b$, $\Delta t_p \sim \Delta t$. In practice this happens when $\bar \kappa \lesssim 1$, notwithstanding extreme values of $\chi$ or $\mu$. Similarly, for $\bar\kappa \gg 1$ one has  $\bar\kappa\Delta t_p \sim \Delta t$. We summarise by the general scaling $\max(1,\bkappa)\Delta t_p \sim \Delta t$. Thus, we see that in general the duration of propagation $\Delta t_p = t_+ - t_-$ is not small compared to the instantaneous visibility $\Delta t$, which justifies taking propagation into account.

The solution Eq. \eqref{eq:propagphipm} exists for any injection time $t_i$ such that $|bt_i| \leq \sqrt{1+a^2}$. This is solved by $t_i \in [t_i{}_-, t_i{}_+]$ with 
\begin{equation}
\label{eq:tipm}
        t_i{}_\pm = \pm \Delta t \frac{\sin\chi|\cos\mu|}{2|\bkappa|}\sqrt{1+a^2} \equiv \pm \frac{\Delta t_i}{2},
\end{equation}
where we have defined the duration of the injection window $\Delta t_i$. At the boundaries of the injection interval $t_+(t_i{}_\pm) = t_-(t_i{}_\pm)$, that is to say the emitting element never crosses the visible segment.   

The burst envelope Eq. \eqref{eq:fpm} is directly obtained by inserting Eq. \ref{eq:propagphipm} into Eq. \eqref{eq:propagzeta} and into Eq. \eqref{eq:r2} in order to obtain an expression similar to Eq. \eqref{eq:fpm},
\begin{equation}
\label{eq:propagfpm}
        f_\pm(t_i) =  f\left(t_\mp(t_i), t_i\right), 
\end{equation}
where the inversion of sign accounts for the fact that in practice later times are associated with a higher altitude and lower frequency.
At this point, the envelope is given as a function of injection time while the observable envelope is a function of arrival time. We derive below the relation between the two and anticipate that, in practice, injection time is a good approximation to arrival time.

It is also convenient to define the characteristic frequency 
\begin{equation}
\label{eq:propagfc}
        f_c(t_i) = f\left(t_c(t_i), t_i\right).
\end{equation} 
We would like to remark that although $t_c$ is the centre of the propagation time interval, $f_c$ is more a characteristic frequency due to its non-linear relation with $t_c$. It is, however, the central frequency of the envelope in the limit of narrow spectral occupancy, that is when Eq. \eqref{eq:propagf} can be linearised. 

\subsubsection{Time of arrival} \label{sec:toa}
The time of arrival is the time measured by the observer when receiving each photon. Assuming vacuum propagation, it is equal to the corresponding emission time $t$ plus the time needed to travel the distance to the observer, that is
\begin{equation}
\label{eq:tate}
t_a = t + \frac{1}{c}\|d\vu - \vec{OM}(t)\|,
\end{equation}
where $d$ is the distance between the centre of the object $O$ and the observer, $\vu$ is the unit vector to the observer, $M$ is the emitting location at $t$, and $\vec{OM}(t)$ is obtained by inserting Eq. \eqref{eq:propagzeta} into Eq. \eqref{eq:OM}. Proper motion of the central object is neglected.

In this section, we consider the effect of subluminal albeit ultra-relativistic propagation by approximating $v/c \simeq 1 - 1/2\Gamma^2$. We expand the right-hand side of Eq. \eqref{eq:tate} up to order 0 in $OM/d$ and drop the constant term $d/c$ for conciseness. Further, we decompose $t = t_i + \delta_i t$ with $\delta_i t = t-t_i$, and expand Eq. \eqref{eq:tate} up to orders $\bigcirc(\Gamma^{-2})$ and $\bigcirc(\Omega^2)$. Expansion is more conveniently carried out using dimensionless spin phases $ (\phi=\omega t,\delta_i\phi= \omega\delta_i t) \sim \Omega$. We obtain
\begin{eqnarray}
\label{eq:tate2}
t_a(t, t_i) & = &  t - \delta_i t\left[1 - \frac{1}{2\Gamma^2} \right.\\
 &  & \left. - \frac{1}{2}\left(\phi^2 \sin^2\chi + \delta_i\phi \phi \bkappa \sin\chi\sin\mu + \frac{\delta_i\phi^2}{3}\bkappa^2 \right) \right] \nonumber.
\end{eqnarray}

We see that to leading order  $t_a = t_i + \bigcirc\left(\Gamma^{-2}, \Omega^2\right)$ which justifies the approximation of Eq. \ref{eq:propagfpm}. It is possible to obtain the exact observable envelope by inserting Eq. \ref{eq:propagphipm} into Eq. \eqref{eq:tate2},
\begin{equation}
\label{eq:tacpm}
        t_a{}_{\{\pm,c\}}(t_i) = t_a\left(t_{\{\mp,c\}}(t_i), t_i\right)
\end{equation}
where the times of arrival $t_a{}_\pm(t_i)$ correspond to the upper and lower frequencies $f_\pm$, Eq. \eqref{eq:propagfpm}, and $t_a{}_c(t_i)$ is a characteristic arrival time which can be associated to a characteristic frequency of the envelope (see Sec. \ref{sec:freqdrift}).  Thus, the envelope in the frequency versus time-of-arrival plane can be drawn in two separate branches $t_i \rightarrow (t_a{}_\pm, f_\pm)$ parametrised by $t_i$. Similarly, a characteristic time of arrival can be derived $t_{ac}(t_i) = t_a\left(t_c(t_i), t_i\right)$

It follows that the total duration of the observed envelope is 
\begin{equation}
\label{eq:Deltata}
        \Delta t_a = |t_a{}\left(t_c(t_i{}_+), t_i{}_+\right) - t_a(t_c(t_i{}_-), t_i{}_-)| = \Delta t_i + \bigcirc\left(\Gamma^{-2}, \Omega^2\right)
\end{equation}
where we recall that at $t_i{}_\pm$ one has $t_c = t_+ = t_-$.

The duration of the crossing of a single emitting element measured by an observer is 
\begin{equation}
\label{eq:deltata}
\delta t_a(t_i) = t_a{}_+(t_i) - t_a{}_-(t_i)  = \Delta t_p \times \bigcirc\left(\Gamma^{-2}, \Omega^2\right)
\end{equation}
where we do not detail the lengthy but straightforward algebra. 

Following the scaling  $\max(1, \bar \kappa) \Delta t_p \sim \Delta t$, Eq. \eqref{eq:propagphipm}, we see that $\max(1, \bar \kappa) \Delta t_p \sim \Delta t_a \simeq \Delta t_i$. It follows that in general $\delta t_a \ll \Delta t_a$ by a factor $\min(1,\bar\kappa^{-1})\times\bigcirc\left(\Gamma^{-2}, \Omega^2\right)$. This has consequences on frequency drifts as we see below.

\subsubsection{Frequency drifts} \label{sec:freqdrift}
Two types of drift must be distinguished: 1) the global drift of the envelope, and 2) the drift due to the propagation of the emitting element. The first is characteristic of downward-drifting sub-bursts that see their characteristic frequency decrease \citep[e.g.][]{hessels_frb_2019}. The second contributes to the apparent slope of a burst in the sense that it is the slope produced by a single point-like emitting element. However, we caution that the slope of an entire burst, resulting from a continuum of emitting elements, is a concept relative to a specific burst model\footnote{For example a two-dimensional Gaussian function which is considered to have a `slope' if its principal axis are rotated with respect to the vertical and horizontal axis \citep[e.g.][]{jahns_frb_2022}.} and cannot be defined in general. 

The characteristic drift of the envelope is obtained noting that $\partial f_c / \partial t_a \simeq \partial f_c / \partial t_i$ (see Eq. \eqref{eq:tate2}),  
\begin{equation}
\label{eq:dfcdta}
        \pderiv{f_c}{t_a} \simeq -  \frac{F}{R_L^\beta}\frac{\beta\omega_i}{\left(x + \omega_i t_i\right)^{\beta+1}},
\end{equation}
where $\omega_i \equiv \omega\crho\left(ab\Delta t/2(1+a^2) - 1\right)$, $F/R_L^{\beta}$ merely normalises the expression to the frequency evaluated at $R_L$, and $f_c$ is defined in Eq. \eqref{eq:propagfc}. The sign of $\omega_i$ determines the sign of the drift. It depends only on the geometrical properties, the spin period, but not on effective opening angle $\Omega$. 

As a useful estimate, one can compute the characteristic drift at the centre of the envelope relative to its central frequency $f_0=F/r^\beta$ (see also Eq. \eqref{eq:f0}). At $t_i\simeq t_a =0$ one gets 
\begin{equation}
\label{eq:fcdot0}
        f_0^{-1}\left.\pderiv{f_c}{t_a}\right|_{0} \simeq -\beta\omega_i/x = -\beta c\frac{\cos\rho}{r} g\left(\epsilon\frac{\bkappa}{\sin\chi}, \sin\mu\right),
\end{equation}
where the second equality results from expanding $\omega_i$ and $g(.,.)$ is a function of purely geometric arguments. The study of $g$ in appendix \ref{ap:fcdot} shows that this function is bounded by $[-1, 1/2]$ such that we are left with 
\begin{equation}
\label{eq:fcdot0bounds}
 -\frac{1}{2}\leq       f_0^{-1} \left.\pderiv{f_c}{t_a}\right|_{0} \left(\beta c\frac{\cos\rho}{r} \right)^{-1} \leq 1.
\end{equation}
Thus the maximum drift amplitude depends only on $\beta, \cos\rho, r$. 
A useful limit is also given by
\begin{equation}
\label{eq:fcdot0limit}
 f_0^{-1} \left.\pderiv{f_c}{t_a}\right|_{0} \underset{\bkappa >>1}{\sim} \beta \omega\frac{\cos\rho}{r\kappa}\frac{\sin\mu}{\epsilon\sin\chi},
\end{equation}
where we have expanded $\bkappa=R_L\kappa$ in order to let appear the proportionality to the spin frequency. 


Second, the propagation drift can be quantified by the average slope 
\begin{equation}
\label{eq:deltafdeltata}
\frac{\delta f}{\delta t_a} =\frac{ f_+(t_i) - f_-(t_i)}{\delta t_a}
\end{equation}
where $f_\pm(t_i)$  is from Eq. \eqref{eq:propagfpm}  and $\delta t_a$ is the single element crossing time from Eq. \eqref{eq:deltata}. 

In most cases, we see that the propagation drift is much larger than the envelope drift, which is consistent with observations \citep[e.g.][]{pleunis_fast_2021, chamma_broad_2022}. Indeed, if the variation of the characteristic frequency is expressed by $\delta f_c = \alpha\delta f$, with $\alpha$ typically of order unity, then  $\partial f_c /\partial t_a \sim \alpha \delta f / \Delta t \ll \delta f /\delta t_a$, which results from Eqs. \eqref{eq:Deltata}-\eqref{eq:deltata}. Following the scaling of Eq. \eqref{eq:deltata} one gets $ \partial f_c/\partial t_a \sim \min(1,\bkappa^{-1})\times\bigcirc\left(\Gamma^{-2},\Omega^{2}\right) \alpha\delta f /\delta t_a$.

\subsection{Polarisation}
\label{sec:PA}
In order to associate a polarisation direction to the envelope we may assume that the rotating-vector model applies \citep{radhakrishnan_magnetic_1969, komesaroff_possible_1970, petri_polarized_2017}. In this model, radiation is polarised in the plane of curvature. This is inspired by the fact that curvature radiation is mostly polarised within in the curvature plane. In our local approach, this means that the polarisation plane is defined by $(\ve_1,\ve_2)$ with normal $\vn_c = \ve_1\times \ve_2 = \ve_3$. Simultaneously, the polarisation vector is orthogonal to the direction of propagation of the electromagnetic wave, that is the direction $\vu$. Thus, the polarisation vector $\vpi$ lies at the intersection of the plane of the sky, orthogonal to $\vu$, and the plane of curvature defined by $\vn_c$, such that
\begin{equation}
\label{eq:pi}
        \vpi = \vn_c \times \vu,
\end{equation}
which is of unit norm. We note that the sign of this definition can be freely chosen since only relative variation of polarisation matter. 

By convention, we define the polarisation angle $\Psi$ with respect to the direction of the spin axis projected onto the plane of the sky, that is $\vom \propto \bot_u \ve_z $. Completed with $\vom\times \vu$ and $\vu$, these three vectors form a complete orthogonal basis.  

Neglecting torsion and curvature variations, the plane of curvature is independent of the position along the streamline such that $\vn_c = \ve_3 = \ve_3|_0$ defined in Eq. \eqref{eq:e30}. Further, one sees that the position angle is fully defined by $\cos\Psi = \vpi \cdot \vom$ and $\sin\Psi = -\vom\times\vu\cdot\vpi$. Thus combining Eq. \eqref{eq:e30}, \eqref{eq:u} and \eqref{eq:pi} one obtains the explicit expressions,
\begin{eqnarray}
        \sin\Psi & = & -\sin\mu\left(\sin^2\chi  + \cos^2\chi \cos\phi\right) - \cos\chi \cos\mu \sin\phi, \\
        \cos\Psi & = & \cos\chi \sin\mu\sin\phi - \cos\mu\cos\phi.
\end{eqnarray}
Combining these two expressions, one readily obtains a formula for $\tan\Psi$ analogous to those of the rotating vector model for pulsars \citep{komesaroff_possible_1970, petri_polarized_2017}. Consistently with our local approximation, we can Taylor expand in $\phi \ll 1$ and obtain the explicit expression 
\begin{equation}
\label{eq:mainpa}
        \Psi(\phi) = \mu + \phi \cos\chi + \frac{1}{2} \phi^2 \sin^2\chi \cos\mu\sin\mu + \bigcirc(\phi^3).
\end{equation}

\newcommand{\rmeas}{\mathrm{RM}}
\newcommand{\pa}{\mathrm{PA}}
\newcommand{\tx}{\tilde x}
\newcommand{\fb}{\bar f^{-1/\beta}}
\newcommand{\aone}{c_{\chi}}
\newcommand{\atwo}{\frac{1}{2}s^{2}_{\chi}c_{\mu}s_{\mu}}

Taking propagation into account, we can express the polarisation angle as a function of the observed frequency and time of arrival. Substituting $\phi$ in Eq. \eqref{eq:mainpa} by the decomposition $\phi = \omega(t_i + \delta_i t)$, where $\delta_i t=t-t_i$ (see Eq. \eqref{eq:tate2}), we can use the map $(t_i, \delta_i t) \rightarrow (t_a, f)$ (Eqs. \eqref{eq:tate2} and \eqref{eq:propagf}) in order to express 
\begin{eqnarray}
\label{eq:PAtaf}
        \Psi(t_a, f) & = & \mu -  \aone \tx  + \atwo \tx^2 \\
        &  & + \omega t_a \left(\aone -\atwo(2 \tx - \omega t_a)\right) \nonumber\\
        &  & + \fb \tx\left(\aone + \atwo\left(2\omega t_a - 2\tx + \tx \fb\right) \right), \nonumber
\end{eqnarray}
where we use the notations $\cos x = c_x, \sin x = s_x$, and $\bar f = f/f_0$ with $f_0=F/r^{\beta}$ (see also Eq. \eqref{eq:f0}), and $\tilde x = r/(R_L\cos\rho)$ with $R_L=c/\omega$ which is the light-cylinder radius and $\cos\rho$ is defined in Eq. \eqref{eq:r2}.
The first line of Eq. \eqref{eq:PAtaf} is constant within the envelope, the second line only depends on $t_a,$ and the third line only depends on $f$ with the exception of one coupling term $t_a\fb$. From Eq. \eqref{eq:mainpa} we see that terms $\propto c_\chi$ are $\bigcirc(\Omega)$ and terms $\propto s_\chi$ are second order and we have suppressed higher-order terms such that effectively $t_a=t_i$ here (see \eqref{eq:tate2}). 

The measured polarisation angle contains a contribution from Faraday rotation, the rotation measure (RM), in addition to the intrinsic polarisation of Eqs. \eqref{eq:mainpa}-\eqref{eq:PAtaf} such that  $\pa = \rmeas \lambda^2 + \Psi$. We see that the leading-order frequency term $\fb \tx\aone$ in Eq. \eqref{eq:PAtaf} exactly mimics RM if $\beta=0.5$. 
In the case of a narrow frequency band, part of the variation of $\pa$ due to Eq. \eqref{eq:PAtaf} can still be absorbed into a fit for RM. To estimate the RM bias $\delta \rmeas$, one can linearise $\pa$ around $f_0$ and see that at leading order, 
\begin{equation}
\label{eq:drm}
        \delta \rmeas \simeq  \frac{f_0^2}{c^2}\frac{r\cos\chi}{2\beta R_L\cos\rho} \simeq 6 \left(\frac{f_0}{1\rm GHz}\right)^{2}\frac{1}{\beta}\frac{r}{R_L}\frac{\cos\chi}{\cos\rho} \mathrm{rad\cdot m^{-2}}.
\end{equation}
This can contribute to the RM fluctuation in narrow-band bursts from repeating sources if different bursts come from different locations. 

One can define a polarisation envelope,
\begin{equation}
\label{eq:psipmc}
\Psi_{\{\pm,c\}}(t_i) = \Psi\left(\omega t_{\{\mp,c\}}(t_i)\right),
\end{equation}
where we compose Eq.\eqref{eq:mainpa} with Eq. \eqref{eq:propagphipm}. This way, the three functions $\Psi_{\{\pm,c\}}$ track PA at the upper, characteristic, and lower frequencies of the burst envelope (Eqs. \eqref{eq:propagfpm} and \eqref{eq:propagfc}). Since at a given $t_a$ variations of Eq. \eqref{eq:PAtaf} are dominated by the term $\fb \tx\aone$, PA varies monotonously between $\Psi_+$ and $\Psi_-$ as a funciton of frequency.

\section{Burst modelling\label{sec:bursts}}
In this section we detail the geometrical effects that intervene between the intrinsic emission process and the observer, and we propose and empirical parametrisation of bursts.

\subsection{Intrinsic spectrum}
Here, we consider the number of photons emitted in the solid angle $\delta O \ll 1$ subtended by the receiver in direction $\vu$ per unit time by the emitting element crossing the fiducial $\zeta =0$ between $t_i$ and $t_i+\dif{t_i}$,
\begin{equation}
\dot N_i \delta O\dif{t_i} \equiv  \frac{\dif{N}}{\dif{t}\dif{t_i}\dif O} \delta O \dif{t_i}.
\end{equation}
In the limit of singular spectrum associated with a given radius, and summing over all emitting elements, we obtain the spectrum per unit time and frequency, 
\begin{equation}
\frac{\dif{N}}{\dif{t}\dif{f}} = \delta O\int \dif{t_i}\, \dot N_i(t,t_i) \delta\left(f - f(t,t_i)\right),
\end{equation}
where $f(t,t_i)$ is defined by Eq. \eqref{eq:propagf}.
This integral yields, 
\begin{equation}
\label{eq:dndtdf}
\frac{\dif{N}}{\dif{t}\dif{f}} = \frac{\dot N_i(t,t_i(t,f))}{\left|\pderiv{f}{t_i}(t,t_i)\right|}\delta O = \dot N_i(t,t_i(t,f)) \frac{F^{1/\beta}}{\beta\cos\rho v f^{1+1/\beta}} \delta O,
\end{equation}
where we made use of the map $(t,t_i) \rightarrow (t, f)$ using Eq. \eqref{eq:propagf} where $r(\zeta_p(t,t_i)) > R_*$ (otherwise the result is trivially 0).

\subsection{Observed spectrum}
In practice, the observed spectrum must be expressed as a function of arrival time $t_a$, Eq. \eqref{eq:tate2}. Using the map $(t, f) \rightarrow (t,\delta_i t) \rightarrow (t_a,f)$, and noticing that frequency $f$ is a function of $\delta_i t$ only, Eq. \eqref{eq:propagf}, we see that the Jacobian of this transformation reduces to $\partial t_a /\partial t$. We can compute the observed spectrum,
\begin{equation}
\label{eq:dndtadf}
\frac{\dif{N}}{\dif{t_a}\dif{f}} = \left.\frac{\dif{N}}{\dif{t}\dif{f}} \pderiv{t_a}{t}\right|_{\delta_i t}^{-1},
\end{equation}
with 
\begin{eqnarray}
\label{eq:dtadte}
\left.\pderiv{t_a}{t}\right|_{\delta_i t}\ & = &  1 + \delta_i \phi\sin^2\chi\left[ \phi \sin\chi + \delta_i\phi \bkappa \sin\mu \right],
\end{eqnarray}
which derives from Eq. \eqref{eq:tate2}. This geometric factor only creates variations of flux of order $\Omega^2$. Notably, there are no caustics. This results from the fact that there is a biunivoque relationship between frequency and position through radius-to-frequency mapping which prevents signal from a range of altitudes to pile up at a given frequency. That could be different if the emitting element had a non-singular spectrum at a given altitude, but this case is out of the scope of this paper. 

\subsection{Injection and angular dependences \label{sec:profiles}}
A convenient case is when the angular dependence of the intrinsic spectrum can be factorised $\dot N_i = A_{\vu}(t,t_i) {\cal\dot N}_i(t,t_i)$ where $A_{\vu}$ is the angular profile that depends on the direction to the observer $\vu$, and ${\cal\dot N}_i$ is the injection temporal profile. Gathering Eqs. \eqref{eq:dndtdf} and \eqref{eq:dndtadf} we get,
\begin{equation}
\label{eq:dndtadf2}
\frac{\dif{N}}{\dif{t_a}\dif{f}} =  \delta O \frac{F^{1/\beta}}{\beta\cos\rho v } \left.\pderiv{t_a}{t}\right|_{\delta_i t}^{-1} f^{-1 - 1/\beta}\,\, A_{\vu}(t,t_i) {\cal\dot N}_i(t,t_i),
\end{equation}
where it is understood that we use the map $(t_a, f) \rightarrow (t, t_i)$.

The time-frequency envelopes studied throughout this work correspond to the edge of an angular profile where the emission stops outside of a cone of angle $\Omega$ around the local tangent while ${\cal\dot N}_i(t,t_i)$ is constant. That can be formally expressed by choosing any angular profile of the form $A_{\vu}(t,t_i) = a_{\vu}(t,t_i) H\left(\Omega - \gamma (t,t_i)\right)$ where $a_{\vu}$ is any smooth function, $H$ is the Heaviside step function ($H(x>0) = 1, H(x \leq 0) =0$), and $\gamma(t,t_i) = \gamma(\zeta_p(t,t_i))$ is the angle between the local tangent of the observer's line of sight as defined in Eq. \eqref{eq:boundarysl}.

\subsection{Pseudo-Gaussian bursts} \label{sec:pseudogaussian}
Several works \citep[e.g.][]{jahns_frb_2022, chamma_broad_2022} have shown that an empirical fit with a multivariate Gaussian parametrised by one width along the time axis, one along the frequency axis, and possibly an extra parameter determining the slope of the burst. We note that in these works the fit can be on the dynamic spectrum itself or on its autocorrelation function. 

Here we propose to use Gaussian angular and injection profiles. The angular profile reads
\begin{equation}
\label{eq:Augauss}
A_{\vu}(t,t_i) = \frac{1}{2\pi\tilde\Omega^2}\exp\left(-\frac{\gamma(t,t_i)^2}{2\tilde\Omega^2}\right)
\end{equation}
where $\gamma^2 \simeq 1-2\cos\gamma = \vu\cdot\ve_1|_0$ is defined in Eq. \eqref{eq:boundarysl}-\eqref{eq:boundarysl2} (see also Sec. \ref{sec:profiles}) and $\tilde\Omega \sim \Omega$ is the characteristic Gaussian width of the profile which is of the same order as the effective angle $\Omega$.  Normalisation is chosen such that $\int A_{\vu} \dif{O} = 1$. For all practical purpose $\dif{O}\simeq \gamma\dif{\gamma} 2\pi$, where $\gamma$ is the colatitude of a local spherical coordinate system,  $2\pi$ arises from integration other azimuth, and integration over $\gamma$ ranges from 0 to infinity as an approximation of the range $[0,\pi]$ taking advantage of the narrowness of the Gaussian. 

In this case the angular profile depends only on $\gamma$. It follows that the envelope defined in Eq. \eqref{eq:fpm} describes an isocontour of $A_{\vu}$. For example, the envelope corresponding to $\Omega = 2\tilde{\Omega}$ delimits the area within which the flux is larger than $\sim 14\%$ of the peak flux at a given frequency, that is the flux it would have had the observer been facing the centre of the beam. 

The injection profile gives the time dependence of the number of photons radiated in all directions per emitting element. It is given by 
\begin{equation}
{\cal\dot N}_i(t,t_i) = \frac{\cal\dot N}{\sigma_i\sqrt{2\pi}}\exp\left(-\frac{(t_i - \bar t_i)^{2}}{2\sigma_i^2}\right),
\end{equation}
where ${\cal\dot N} = \int {\cal\dot N}_i \dif{t_i}$ is the total number of photons radiated per unit time (in all directions) by all emitting elements at a given instant $t$. The most powerful element is injected at the mean time $t_i = \bar t_i$ and the characteristic duration is $\sigma_i$, that is approximately $68\%$ of all emitted photons are emitted by elements injected between $[\bar t_i -\sigma_i, \bar t_i + \sigma_i]$. Note that this does not in general represent $68\%$ of photons that are seen because this is modulated by the visibility due to the angular profile. Note that for simplicity this profile has no dependence on $t$, which means that an emitting element emits constantly at all times inside and outside the visible region. Equivalently, this means that the same number of photons is emitted at all frequencies. If the visible region is small enough or the frequency dependence is soft enough such that little variations occurs across the visible region then this is a good approximation.

One can see that once composed with the map  $(t_a, f) \rightarrow (t, t_i)$ and injected into Eq. \eqref{eq:dndtadf2} these profiles do not produce a Gaussian burst, neither in time nor in frequency. However, we have used Gaussian functions to describe the two intrinsic physical profiles, hence the qualification of pseudo-Gaussian bursts.

\section{Applications to simple magnetospheric geometries \label{sec:applications}}

In the local description presented in Sec. \ref{sec:model}, burst envelopes depend on 9 parameters, namely $\left(\omega,\chi, \beta, F, \Omega, \mu, \cos\rho, \kappa, r\right)$. The first 3 are global parameters, common to all bursts from the same source: they describe the spin rate of the source, $\omega$, the inclination of the spin axis with respect to the observer, $\chi$, and the radius-to-frequency mapping power-law index and normalisation of the emission process, $\beta$ and $F$. On the other hand, the following 5 parameters are local parameters a priori independent from an envelope to another. In practice, an envelope corresponds to a close sequence of bursts or sub-bursts. The opening angle $\Omega$ encodes our ignorance about the emission process and exact distribution of emitting plasma (see Sec. \ref{sec:bundlereduc}), while $\left( \mu, \cos\rho, \kappa, r\right)$ describe the local geometry. These local geometrical parameters are however related to one another once a global geometry is chosen. Formally, a global geometry is described by a vector field $\vec{v}(\vr, \vec{p})$ (for instance, a velocity field) that generates streamlines as a function of global parameters $\vec{p}$ (which may include $\omega,\chi, \beta$). Local geometrical parameters are thus derived from $\vec{v}$ and become functions of position $\vr$ and global parameters $\vec{p}$. As per Eq. \ref{eq:chi}, there is a relation between $\chi$ and position, which removes a degree of freedom on the position of the emission region for each envelope. Thus, only 3 local parameters remain: 2 for position and $\Omega$.

When a model of bursts is specified, additional local parameters must be taken into account for each individual (sub-)burst. In the case of pseudo-Gaussian bursts of Sec. \ref{sec:pseudogaussian}, these are $(\bar t_i, \sigma_i, \cal\dot N)$ which characterise the mean injection time, burst duration and photon flux, respectively. If one considers bursts rather than envelopes, the Gaussian width of the emission profile, $\tilde \Omega$, replaces $\Omega$. 

In this work, we study two simple magnetospheric geometries. We assume that the emitting plasma approximately follows magnetic field lines, which consequently are streamlines. This is expected in acceleration regions where the force-free approximation does not apply and the magnetic field dominates, such as gaps along open magnetic-field lines in pulsar magnetospheres \cite[e.g.][ for a review]{petri_theory_2016}. Thus, such geometries may apply to FRB models that consider pulsar-like mechanisms along open magnetic-field lines, propagation of Alfv\'en waves, or accelerated particle bunches. This is however not valid if energy is conveyed by fast magnetosonic waves or if the magnetosphere is massively deformed during the burst, for instance. For convenience, we choose the streamlines and the magnetic field to share the same orientation. 

In a neutron-star magnetosphere, pulsar or magnetar, inner currents are expected to dominate close to the star such that vacuum magnetic-field solutions are a good approximation. In addition, co-rotation velocities are small compared to the speed of light such that static solutions can be used. In Sec. \ref{sec:dipole}, we consider the simplest of these vacuum static solutions, that is a static dipolar magnetic field the source of which is at the centre of star. 
We show that this type of model can be used to produce bursts reminiscing of one-off events. 

In Sec. \ref{sec:dipoletoro}, we consider a magnetic field resulting from the sum of a dipole and a strong toroidal component. The magnetic field thus constructed does not globally represent a physical magnetosphere. However, it reproduces local conditions and we use this model to explore the effects of an added toroidal component. The motivation for adding a toroidal component is twofold. First, we argue that within our model it appears as the easiest and perhaps the only way to generically produce narrowband downward-drifting patterns akin to what is observed from repeating FRB sources. Second, there are strong theoretical as well as observational arguments in favour of strong multipolar or toroidal components in the magnetic field of magnetars, either from internal currents \citep[e.g.][]{barrere_new_2022} or from magnetospheric currents \citep[e.g.][]{thompson_electrodynamics_2002, beloborodov_untwisting_2009}. Observational arguments for additional components revolve around the detection of magnetic field intensities much larger than the dipolar component inferred from the stellar spindown. Observations can either be x-ray spectral features directly associated with magnetic intensity \citep[e.g.][]{rodriguez_castillo_outburst_2016}, or estimates of the internal toroidal magnetic field necessary to produce the inferred free precession of the star \citep[e.g.][]{makishima_discovery_2021}. In all cases, the estimated non-dipolar fields can be more than two orders of magnitude larger than the inferred dipolar component. 

We would like to note that in rotating magnetospheres, a toroidal component is generically generated by rotation effects, leading to spiralling field lines at distances of the order of or larger than the light-cylinder radius. The simplest example is given by the split-monopole solution \citep{michel_rotating_1973}. However, in this case the field becomes significantly toroidal only at large distances where effects of a relativistic co-rotation velocity must be taken into account and therefore the present model cannot be applied as is. Moreover, the effect of the toroidal component in the split-monopole model is neutralised by the effect of relativistic composition of velocities. Indeed, one can see that in the split-monopole case, particles follow the magnetic field lines in the co-rotating frame, but they do in fact follow a purely radial trajectory in the inertial frame. As a result an emitting element propagating towards the observer remains visible virtually indefinitely and an envelope cannot be defined.

In this section, all frequencies are normalised by the frequency at the centre of the emission region, 
\begin{equation}
\label{eq:f0}
f_0 \equiv Fr(\zeta=0)^{-\beta} = Fr^{-\beta}
.\end{equation}

For both geometries we assume an infinite Lorentz factor $\Gamma$ in order to simplify the discussion. It is enough to note that the main effect of a finite Lorentz factor is to slow down the propagation drift whenever $\Gamma \lesssim \Omega$ (see Sec. \ref{sec:freqdrift}).

\subsection{Dipolar polar cap\label{sec:dipole}}
Below, we present an approximate treatment restricted to magnetic colatitudes in the vicinity of the last open field lines. This allows us to considerably simplify the algebra and neglect propagations effects as well while retaining key elements. If reasonably accurate, we nonetheless compute numerically Figs. \ref{fig:envdip} and \ref{fig:burstJ1935} in the subsequent sections so that no approximation is made in the figures. 

\subsubsection{Analytical approximation}
For simplicity, we restrict analytical derivations to the case of open magnetic-field lines which at low altitudes constitute a narrow bundle of lines around the magnetic poles and are characterised by a magnetic colatitude $\alpha \leq \alpha_{\rm o}(r) = \sqrt{r/R_L}$ (see appendix \ref{ap:dipolegeom}). This particular region is interesting because its open topology provides favourable condition for particle acceleration that is key in pulsar magnetosphere models \cite[e.g.][]{petri_theory_2016}. Besides, we see that its phenomenology is very diverse.

The magnetic field in the limit of low magnetic colatitude $\alpha \ll 1$ can be expressed by
\begin{equation}
\label{eq:Bdip}
\vB= \frac{B_L}{x^3} \left(3\ve_r' -\vn\right),
\end{equation}
where $B_L$ gives the scale of the magnetic field at the light cylinder $R_L$, $\ve_r'=\vr/r$ is the radial unit vector in the rotating frame $F'$, and $x= r/R_L$. The magnetic dipole moment is in the direction of the unit vector $\vn = (\sin i,0,\cos i)_{F'}$, inclined by an angle $i$ relative to the spin axis of the star. 

We give the detailed analytic expressions of the relevant derived quantities in appendix \ref{apsec:dipole}. In particular we see that the reduced curvature $\bar \kappa = \kappa R_L \gg 1 $ as long as $\alpha \gg r/R_L$, which is true in most of the region above the polar cap except where $|\alpha| \ll \alpha_{\rm o}$. We show in appendix \ref{apsec:dipole} that this implies a small propagation time, that is $\Delta t_p \ll \Delta t$ in Eq. \eqref{eq:propagphipm}, and that the limit of instantaneous propagation applies. This limit allows us to derive simple expressions that are useful to apprehend the behaviour of these envelopes. We develop this approach in appendices \ref{ap:instenv} and \ref{apsec:dipole} and report below only the key results. Thus, for the rest of this subsection we assume $\bar\kappa \gg 1$ and neglect propagation such that $t = t_a = t_i$, except otherwise stated. 

We derive the instantaneous visibility envelope, Eq. \eqref{eq:fpm} and Eq. \eqref{apeq:fpm},
\begin{equation}
\label{eq:dipoleenv}
        f_\pm = f_0\left[\max\left(R_*/r, 1 - y(x,\alpha) \left(\sin i \frac{\delta\varphi}{|\delta\theta|}\bar\phi \mp  \sqrt{1-\bar\phi}\right)\right)\right]^{-\beta},
\end{equation}
where $f_0$ is defined in Eq. \eqref{eq:f0}, and we have introduced the scaling $y(x,\alpha) = 4\Omega /3\alpha $. The longitude and colatitude shifts $\delta\phi, \delta \theta$ parametrise the location relative to the axis of the magnetic pole and are related to the magnetic colatitude $\alpha$ by Eq. \eqref{apeq:alpha}.

The value of $y$  governs spectral occupancy. Where $y \gtrsim 1$, that is $\alpha \lesssim \Omega$, one expects a broad spectral occupancy whenever $\sin i|\delta\phi| \gtrsim |\delta\theta|$. By broad, we mean that $\Delta f /f_0 > 1$. In any case, due to the singularity in $\delta \theta$ there is always a region of broadband emission, although we caution that our leading-order approximation breaks down at $\delta \theta =0$. 

Another noteworthy property of low-altitude dipolar geometry is the fact that frequency can drift up or down, depending on the location with respect to the magnetic axis. Indeed, it is enough to calculate Eq. \eqref{eq:fcdotfc} at $\bar\phi=0$,
\begin{equation}
\label{eq:fcdotdipole}
\left.\frac{\dot f_c}{f_c}\right|_0 = \frac{4}{3}\beta\omega \sin^2i \frac{\delta\varphi}{\alpha^2},
\end{equation}
where the dot denotes time derivative and we note that $f_c|_0 = f_0$. One sees that the sign of the drift is the sign of $\delta\varphi$ meaning that the drift direction depends on whether the line of sight is before or after the magnetic pole along the (spin) azimuthal direction.

The duration of the envelope is given by Eq. \eqref{eq:dt} which yields,
\begin{equation}
\label{eq:dtdipole}
        \frac{\Delta t}{P} = 2\frac{\alpha}{\sin i |\delta \theta|} \frac{\Omega}{2\pi},
\end{equation}
where one sees that longer envelopes are obtained for smaller colatitude shift $\delta \theta$ (for a fixed magnetic colatitude $\alpha$). Similarly to the sign of frequency drift, this is due to the plane of curvature of these field lines being almost parallel to the plane tangent to the cone generated by the line of sight in the rotating frame ($\mu$ close to $\pm 90^\circ$).

If we totally neglect propagation, then the propagation drift of Eq.  \eqref{eq:deltafdeltata} is simply $\delta f /\delta t =\infty$. Going one step further, we have seen in Sec. \ref{sec:propagation} that the propagation drift is typically $\delta f /\delta t \sim \max(1,\bar\kappa)\times\bigcirc\left(\Gamma^{2}, \Omega^{-2}\right)\dot f_c$.

\subsubsection{Envelopes \label{sec:dipoleenv}}
The presence of singularities in $\delta\theta,\delta\varphi$, and $\alpha$ appearing in Eqs. \eqref{eq:dipoleenv}-\eqref{eq:dtdipole} (themselves resulting from singularities in Eqs.  \eqref{apeq:kappadipole}-\eqref{apeq:hdipole}) results in extreme variations of the envelopes around the magnetic pole. This explains the diversity of morphologies that can be found in the polar-cap region alone. On the other hand, this means that field-line properties can vary very quickly close enough to the pole and therefore that our approximations, which require a certain smoothness in the variations of geometrical properties, break down at the centre of the region.

\begin{figure*}
        \centering
        \includegraphics[width=1\textwidth]{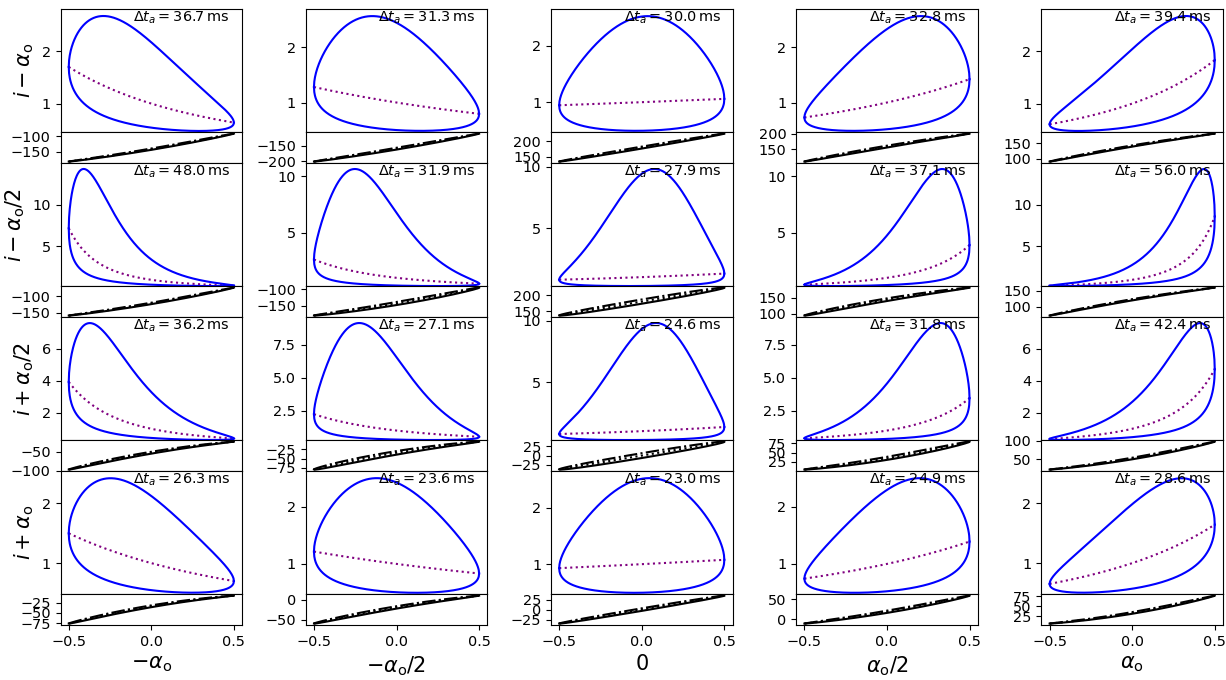} 
        \caption{Burst envelopes (upper panels) and PA (lower panels) for various colatitudes (vertically) and longitudes (horizontally) centred on the magnetic axis in the vicinity of the last open field lines (magnetic colatitude $\alpha \sim \alpha_{\rm o}$) at altitude $r = 100 R_*$ and for magnetic inclination $i=45^\circ$.  
        Characteristic burst frequencies $f_c$ (Eq. \eqref{eq:propagfc}) are shown as dashed purple lines, while blue contours show $f_\pm$ (see Eq. \eqref{eq:propagfpm}). Frequencies (y axis of upper plots) are represented in units of $f_0 = F/r^\beta$. 
        Characteristic PA, $\Psi_c$, are dotted black lines, $\Psi_+$ are solid black lines and $\Psi_-$ dash-dotted lines (see Eq. \eqref{eq:psipmc}). PA are in degrees.
        The following fiducial values are assumed: $P=3245$ms, $\Omega=\alpha_{\rm o}/5\simeq 0.09$ rad, $\beta=3$.  }
        \label{fig:envdip}
\end{figure*}

 In order to consistently restrict the discussion to the polar-cap region, one must choose $\Omega \lesssim 2\alpha_{\rm o}$. This can be seen from the fact that the field lines being quasi-radial in that region, the typical opening angle of a bundle cannot be much wider than $2\alpha_{\rm o}$. Note that this only follows from our arbitrary choice of restricting the discussion to the polar-cap region.
In Fig. \ref{fig:envdip}, we choose $\Omega=\alpha_{\rm o}/5$, $P=3.245$s and $r=100R_*$ giving $r/R_L \ll 1$ and $\Delta t \sim 30 - 50$ ms. These values allow us to remain in the regime where burst-duration is trigger-limited (see Sec. \ref{sec:bundlereduc}) and co-rotation is non-relativistic. The period is that of magnetar SGR 1935+2154 responsible for the Galactic FRB 200428 \citep{andersen_bright_2020, bochenek_fast_2020}. The choice of this particular value, albeit arbitrary, will be useful in the discussion.

We choose a fiducial radius-to-frequency-mapping $\beta=3$. This choice results in wide spectral occupancy everywhere on the grid of coordinates of Fig. \ref{fig:envdip} for the given set of parameters. This value can also take a physical meaning in case emission frequency scales linearly with the local magnetic strength. 
We stress that, as discussed above, spectral occupancy diverges in the innermost region of the polar cap which allows for broadband emission regardless of the value of $\beta$.

 Thus, broadband emission is easily obtained in at least part of the polar-cap region (even with lower values of $\beta$). Frequency drift can go either up or down depending on the hemisphere around the magnetic pole (see Fig. \ref{fig:envdip}) but independently of altitude. The large bandwidth is a consequence of the field lines being radial and curvature begin small (and even cancels at the pole). 
 
 Frequency drift, Eq. \eqref{eq:fcdotdipole}, increases in magnitude with azimuthal shift $|\delta\varphi|$. At altitude $r=100R_*$ of our example, the drift is weakly but visibly affected by propagation effects as can be seen in Fig. \ref{fig:envdip} by the fact that the drift rate does not cancel at $\delta\varphi=0$. The magnitude of the effect is better obtained numerically, giving $\dot f_c/f_0 \sim 55 \rm s^{-1}$ at $|\delta\varphi| \sim \alpha \sim \alpha_o$. This is comparable to the approximate rate of Eq. \eqref{eq:fcdotdipole}, $f_0^{-1}|\partial f_c/\partial t_a| \simeq 44 \; (\beta/3)(\sin i/\sin 45°)^2 (3.2\;\mathrm{s}/P)^{1/2} \rm s^{-1}$.  As shown in Fig. \ref{fig:burstJ1935}, this rate appears compatible with the two bursts of the Galactic magnetar FRB 200428 observed by CHIME/FRB \citep{andersen_bright_2020}. However, it is also clear from Fig. \ref{fig:burstJ1935} that one cannot measure the characteristic frequency drift from the data due to the broadband nature of the bursts. However, the upper and lower frequencies of the bursts constrain the envelope. Visual inspection of the few one-off events showing upward frequency drift in the CHIME/FRB catalogue \citep{pleunis_fast_2021} suggest that their drift rates could be similar. As cautioned in \citet{pleunis_fast_2021} and above in this paragraph, their broadband nature hinders the determination of their central frequency\footnote{\citet{pleunis_fast_2021} define empirically a central frequency, while we defined in this paper a characteristic frequency which is in general not equivalent, especially for broadband bursts. However, their qualitative behaviour concerning this paragraph is similar.} drift due to the limited bandwidth of the instrument. Thus, we suggest that if the lower or upper edge of the occupied frequency band is within the instrument range, it may be less ambiguous to measure the drift of the edge and compare it to the envelope boundaries, Eq. \eqref{eq:dipoleenv}, rather than trying to determine the characteristic (or central) frequency drift. 

Remarkably, this is the situation of the two components of FRB 200428 observed by CHIME/FRB \citep{andersen_bright_2020} and  associated with SGR 1935+2154. These bursts are separated by $\sim 29$ ms and the first one occupies a frequency band $f < f_+^{(a)} \simeq 650$ MHz while the second one occupies $f> f_-^{(b)}\sim 450$ MHz. The minimum frequency of the first burst is outside the band of the instrument, but the second burst was also observed by STARE2 between $1280$ and $1468$ MHz, so we may assume that $f_+^{(b)} \geq 1468$ MHz. We see from these properties that the two components of FRB 200428 can be bounded by an envelope similar to those at $(\alpha_{\rm o}, i \pm \alpha_{\rm o}/2)$ in Fig. \ref{fig:envdip} (see also Sec. \ref{sec:burstexdp} and Fig. \ref{fig:burstJ1935}). 

PA is changing mostly linearly and at a relatively slow rate of $\sim 2^\circ/\rm ms$ such that even if the linear drift was not absorbed in RM (Sec. \ref{sec:PA}) it would not easily be detected over the duration of a single burst. On the other hand, its central value reverses sign across the magnetic pole. This comes from the fact that $\Psi(\phi=0, \delta\theta, \pm\delta\varphi) = \pm \mu(\delta\theta,\delta\varphi)$ as can be seen from Eq. \eqref{eq:mainpa} and Eqs. \eqref{apeq:dipcosmu}-\eqref{apeq:dipsinmu}. Thus, for a given line of sight a broad range of PA can be observed depending on the spin phase at the moment of the burst.  
Regarding the case of FRB 200428, we note that no variation of PA was measured by CHIME/FRB between the two components of FRB 200428 within an uncertainty of $30^\circ$ \citep{andersen_bright_2020}, which is compatible with envelopes in Fig. \ref{fig:envdip}.

\subsubsection{Burst example} \label{sec:burstexdp}
\begin{figure}
        \centering
        \includegraphics[width=0.5\textwidth]{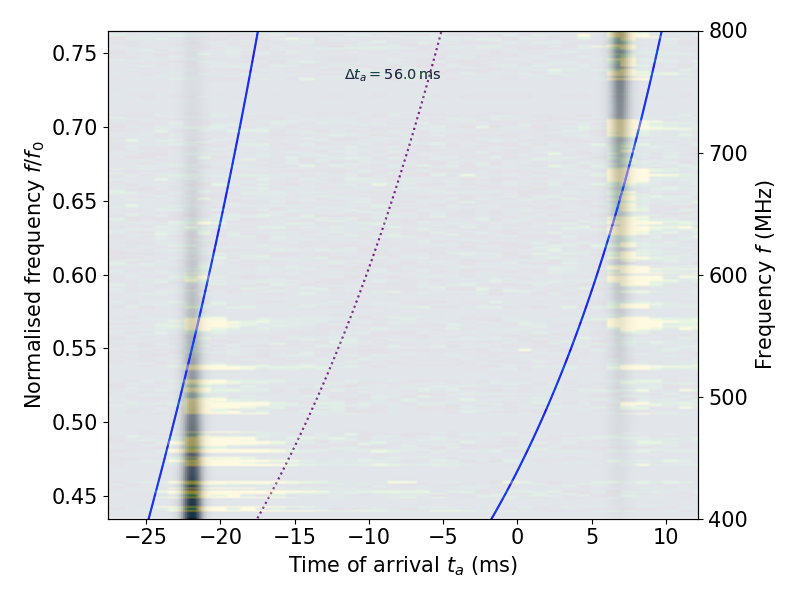}
        \caption{Frequency versus time of arrival plot of two bursts (grey scale) in polar-cap geometry overlaid with the two bursts observed from the Galactic magnetar SGR J1935+2154 by CHIME/FRB \citep[extracted from Fig. 1 of ][]{andersen_bright_2020}. As in the CHIME/FRB observation, the two bursts are separated by 29ms, and the frequency range has been restricted to $f/f_0 \in[0.4, 0.8]$, which can be immediately scaled to the $400-800$MHz bandwidth of the telescope (right-hand-side vertical axis). The second burst has a 25\% larger photon rate $\cal \dot N$, and they have the same Gaussian width $\sigma_i=0.5$ms and Gaussian opening angle $\tilde\Omega=\alpha_{\rm o}/10$. Other fiducial parameters are the same as in Fig. \ref{fig:envdip} for the envelope at colatitude $i-\alpha_{\rm o}/2$ and longitude $\alpha_{\rm o}$: $P=3245$ms (period of SGR J1935+2154), $\Omega=\alpha_{\rm o}/5\simeq 0.020$ rad, $\beta=3$. Line styles for the envelope are as in Fig. \ref{fig:envdip}.}
        \label{fig:burstJ1935}
\end{figure}
The two pseudo-Gaussian bursts (Sec. \ref{sec:pseudogaussian}) in Fig. \ref{fig:burstJ1935} are generated using parameters aimed at mimicking the two bursts seen by CHIME/FRB from the Galactic magnetar SGR J1935+2154 \citep[Fig. 1 in][]{andersen_bright_2020}. In particular, the two bursts are separated by 29 ms and the spin period is that of SGR J1935+2154. The envelope (in blue) is a zoom-in of the envelope at colatitude $i-\alpha_{\rm o}/2$ and longitude $\alpha_{\rm o}$ in Fig. \ref{fig:envdip} with $\Omega = 2\tilde\Omega = \alpha_{\rm o}/5$ making it the contour where the flux is $\sim 14\%$ of its maximum value at this frequency. The frequency range can be normalised to the telescope's bandwidth of $400-800$MHz in order to make these two bursts analogous to the observation both in time and spectral occupancy. It is noteworthy that the second burst is also compatible with STARE2's observation since the upper frequency of the envelope lies well above 1468MHz once the frequency axis is scaled to fit CHIME's observation. Interestingly, the envelope that `fits' the properties of the two observed bursts corresponds to a field line at the edge of the open-field-line region. 

Here we chose a photon rate $\cal \dot N$ that is 25\% larger for the second burst in order to improve contrast. A more realistic treatment would also include differentiated durations $\sigma_i$ between the two bursts as well as convolution by a scattering function in order to account for scattering by the interstellar medium \citep[e.g.][]{suresh_48_2021}. 

The bursts are broadband (compared to the observing bandwidth), and we can see that the envelope drift is much better estimated by looking at the upper frequency cut-off (first burst) and low-frequency threshold (second burst) than at the characteristic frequency which is not measurable for either of them.

As expected from the discussion in Sec. \ref{sec:dipoleenv} about dipolar polar-cap geometry, the propagation drift (that is the slope within each burst) is too steep to be visible by eye on these plots.

\newcommand{\rmax}{r_\max}

\subsection{Effect of a strong toroidal field \label{sec:dipoletoro}}

In this section we investigate the generic consequences of the presence of a strong toroidal component wrapping around the spin axis. 
To this end, we consider a magnetic field of the form
\newcommand{\Btoro}{B_{\rm toro}}
\begin{equation}
        \vB = \vB_{\rm dip} + \Btoro \ve_\varphi,
\end{equation}
where $\vB_{\rm dip}$ is a dipolar field of the form given in Eq. \eqref{eq:Bdip}, $\ve_\varphi$ is the azimuthal unit vector with respect to the spin axis, and $\Btoro$ is the amplitude of the toroidal component. It should be reminded that within our convention co-rotation velocity is along $+\ve_\varphi$. 

\begin{figure}
        \centering
        \includegraphics[width=0.5\textwidth]{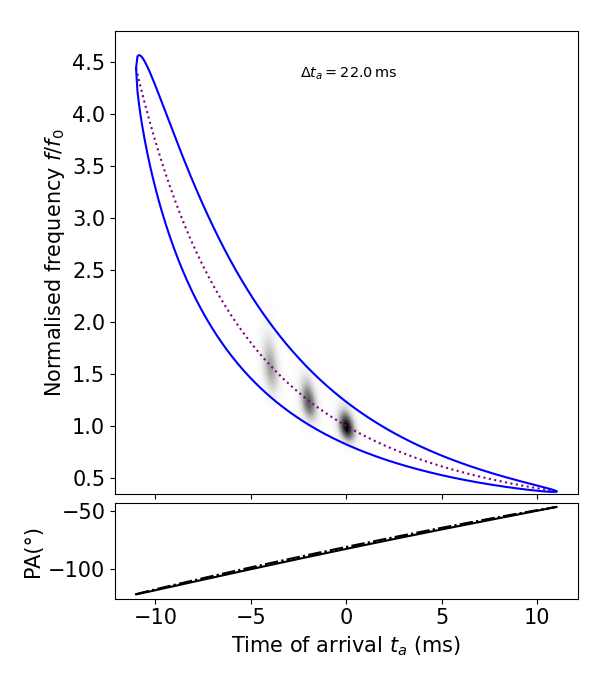} 
        \caption{Repeater-like bursts, envelope, and polarisation. Top: Frequency versus time of arrival plot of a sequence of three bursts (grey scale) and an envelope. Bottom: PA envelope. The geometry is dipolar with an added toroidal component which intensity if 50\% that of the poloidal component. The location is at the vertical of the magnetic pole of the dipole at altitude $r=100R_*$, as in Fig. \ref{fig:envdip}. Other parameters are also identical to Fig. \ref{fig:envdip}, in particular $\Omega\simeq 0.09$rad  and $\beta = 3$, except the spin period which is $P = 250$ms (see main text). The three bursts have identical photon rates $\cal \dot N$, identical Gaussian width $\sigma_i=0.3$ms and Gaussian opening angle $\tilde\Omega=\Omega/2$. Line styles for the envelope and PA curves are as in Fig. \ref{fig:envdip}.}
        \label{fig:bursttoro}
\end{figure}

In Fig. \ref{fig:bursttoro} we show an example of envelope where $\Btoro = -0.5 \|\vB_{\rm dip}\|$. The location has been chosen at the vertical of the magnetic pole such that the dipolar component is purely radial in order to simplify interpretation. We caution that the altitude $r=100R_* \simeq 0.1R_L$, chosen to be identical to Fig. \ref{fig:envdip}, is in the upper range of validity of the co-rotation hypothesis and that as a result an error of order $\sim r/R_L=10\%$ is to be expected. Figure \ref{fig:bursttoro} shows what can be generically observed whenever the toroidal field is strong enough: the envelope drifts monotonously and tends to be narrow-band. The drift goes downwards if the streamlines go against the rotation (reminding that signs are chosen such that the stream is in the direction of $\vB$), and upwards otherwise. This is reminiscing of the properties often seen in burst from repeating sources, and the properties of the envelope and bursts in Fig. \ref{fig:bursttoro} have been chosen in order to produce a pattern similar to bursts from FRB 121102 published in \citet{hessels_frb_2019}. We now detail further the properties. 

We see in Eq. \eqref{eq:dfcdta} that the sign of the envelope drift is the sign of $-\omega_i$. A straightforward expansion of $\omega_i$ further shows that whenever the emitting element is oriented outwards, that is $\cos\rho > 0$, 
\begin{equation}
\label{eq:signfcdot}
        \mathrm{sign}\left(\pderiv{f_c}{t_a}\right) = \mathrm{sign}\left(\cos\rho\left[1 + \frac{ \bar\kappa\sin\mu}{\sin\chi}\right]\right),
\end{equation}
where as before $\chi$ is the observer's colatitude, $\bkappa = \kappa R_L$ is the reduced curvature and $\mu$ is the curvature-plane angle. 

A purely toroidal field has circular field lines of radius $r$ centred on the spin axis $\ve_z$. It follows trivially that $\bkappa = R_L/r$ such that $\bkappa \gg 1$ at low altitudes compared to $R_L$, and $\chi=90^{\circ}$. We have from Eq. \eqref{eq:sinmu} that $\sin \mu = (\ve_1|_0\times\ve_2|_0)\cdot \ve_z/\sin\chi$ where $\ve_1|_0 \propto \vB$ is the local tangent, $\ve_2|_0$ points towards the centre of curvature, and $\ve_z$ is the spin axis. It follows that $\sin\mu = \mathrm{sign}(\Btoro)$. Thus, the sign of $\Btoro$ determines the sign of the expression between square brackets in Eq. \eqref{eq:signfcdot}. On the other hand $\cos\rho= 0$ such that the envelope is in fact empty for a purely toroidal field. Nonetheless, it follows from continuity that the sign of the drift associated with a more complex field is still determined by $\Btoro$ as long as it is strong enough and the plasma propagates outwards, that is $\cos\rho>0$. This property can in fact be used to qualify the strength of the added toroidal component: it is strong compared to the other components as long as the sign of the drift associated with it is preserved.

In practice, a numerical exploration of locations in the magnetosphere with the parameters of Fig. \ref{fig:bursttoro} shows that a relative intensity of the added toroidal component of 50\% is sufficient to have a uniform drift orientation everywhere in the vicinity of the dipolar polar cap, and almost everywhere in the magnetosphere. 

In the case of Fig. \ref{fig:bursttoro}, we have $\bar\kappa \simeq 5.6$ and $\sin\mu \simeq -0.99$. This leads to a rate of envelope drift of $f_0^{-1}\partial f_c/\partial t_a \sim 110 \rm s^{-1}$ at $t_i \simeq t_a =0$, Eq. \eqref{eq:dfcdta}.  This is to be compared to the rate of $-150 \rm s^{-1}$ reported for FRB121102 \citep{2019ApJ...882L..18J} or the rates of $\sim-4$ to $\sim-40 \rm s^{-1}$ reported for FRB180916 and a few other FRBs in \citet{andersen_chimefrb_2019}. 

The propagation drift is  $f_0^{-1} \delta f /\delta t_a\sim 10^5 \rm s^{-1}$ at $t_i =0$, Eq. \eqref{eq:deltafdeltata}, following roughly the scaling $ \delta f /\delta t_a\sim \bkappa \Omega^{-2} \partial f_c/\partial t_a $ for $\bkappa \gg 1$. Thus, the propagation drift is much too large to be visible by eye in Fig. \ref{fig:bursttoro} and the slight `slope' that can be seen is rather due to the non-Gaussianity of pseudo-Gaussian bursts. It could be interesting to see how this asymmetry may be effectively fitted as a slope by Gaussian models \citep[e.g.][]{chamma_broad_2022, jahns_frb_2022}, but we differ this to further work. 

The frequency bandwidth associated with a purely toroidal field is zero. Indeed, altitude is constant along field lines which leads to a null bandwidth within the hypothesis of radius-to-frequency mapping. In Eq. \eqref{eq:signfcdot} this translates into $\cos\rho=0$.
Again, by continuity from the purely toroidal case one sees that a strong toroidal component tends to produce narrow bandwidths. In Fig. \ref{fig:bursttoro}, we have $ \delta f/f_0 \simeq 0.4$ at $t_i=0$, similar to what can be seen from known repeaters. 

In the inner magnetosphere where $\bar \kappa \gg 1$ the duration of the envelope is $\Delta t_a = \Delta t + \bigcirc(\Gamma^{-2}, \Omega^2) + \bigcirc(\bar\kappa^{-1})$, Eqs. \eqref{eq:Deltata} and \eqref{eq:tipm}. Reminding ourselves that $\Delta t = 2P\Omega/(\sin\chi|\cos\mu|)$, Eq. \eqref{eq:dt}, we see that all else being equal the addition of a toroidal component lengthens the duration of the envelope as $\cos\mu \rightarrow 0$. This is what has allowed us to reduce the spin period to only 250 ms in Fig. \ref{fig:bursttoro} compared to Fig. \ref{fig:burstJ1935} while retaining a similar envelope duration. On the other hand, this shorter spin period is necessary to obtain a drift rate as large as observed in repeating FRBs (see above). A similar drift rate could be achieved by reducing altitude instead of spin period, as can be deduced from Eq. \eqref{eq:fcdot0limit}, however, at the cost of a different frequency normalisation $F$ in order to remain within the same frequency range. Equations \eqref{eq:fcdot0}-\eqref{eq:fcdot0limit} show that a change of line of sight $\chi$, or of an intrinsic parameter among $\mu, \rho, \kappa, \beta$ can also affect the drift rate. On the other hand, a reduction of $\Omega$ would not have this effect, and would also further reduce the bandwidth. Thus, if a strong toroidal component is responsible for downward drifting bursts in repeaters, this example suggests that it must be associated with faster rotation in order to match observations, all else being equal. 

Concerning PA, the case of Fig. \ref{fig:bursttoro} shows a mostly linear drift  and little difference between the higher and lower frequency parts of the envelope that are very similar to the PA envelopes discussed in Sec. \ref{sec:dipoleenv}. The conclusions are thus the same: even if not absorbed in RM, Sec. \ref{sec:PA}, the variation of PA is probably too slow to be detected within a single burst. 

\section{Discussion \label{sec:discussion}}

We consider that FRBs are produced on a narrow bundle of streamlines by a perturbation propagating outwards from the surface of a neutron star along those lines. This can be the case if the bundles are excited by star quakes which occur randomly at the surface of the star as in models of magnetar outbursts. Since no fast periodicity has been evidenced so far in repeaters we further consider that emission can be produced in different regions of the magnetosphere in an apparently random manner. For definiteness we assume that streamlines follows magnetic-field lines as calculated in this paper.

The qualitative properties of the model are quite independent from the specific global model of the streamline geometry. This entails from the fact that the model is local, and observed properties essentially result from the magnitude and orientation of the local curvature, as well as the effective opening angle. The main non-local parameter is certainly the spin frequency of the rotating frame. If one considers the close environment of a neutron star, as we do here, the orders of magnitude of some of these quantities are quite constrained. The spin period between a few milliseconds and a few seconds, the radius of curvature between the stellar radius and the light-cylinder radius except in some very localised places such as magnetic poles where curvature can tend to zero. The orientation of the curvature plane is a priori unconstrained but plays a major role in the direction and magnitude of envelope drifting, Sec. \ref{sec:dipoletoro}. The effective opening angle mostly controls the extent of envelopes, both in terms of duration and bandwidth. If one considers open field line, then its magnitude can be scaled to the polar cap. The remaining parameter, the radius-to-frequency-mapping index, is independent from the chosen geometry. In Sec. \ref{sec:applications}, we have presented a set of local envelopes which covers a large part of the geometrical parameters space in terms of curvature magnitude and orientation. 

Locality also implies that several narrowband bursts cannot be seen simultaneously in different frequency bands unless several bundles are excited simultaneously. Indeed, for simple convex geometries the observer only sees one segment of the streamline at a time which can only result in contiguous spectral occupancy. This is consistent with observations of repeaters thus far, so we only consider the case of a single excited bundle. 

We have shown in Sec. \ref{sec:dipoletoro} that narrowband downward drifting bursts can be generically produced if streamlines have a strong toroidal component wrapping around the spin axis in the direction opposite to the co-rotation velocity. 
On the other hand, we have seen in Sec. \ref{sec:dipole} that broadband events, drifting either upwards or downwards, can be produced in the polar-cap region of a dipolar field. This is compatible with the fact that all three one-off events with apparent upward drifts in the CHIME/FRB catalogue are also broadband \citep{pleunis_fast_2021}, as well as the two components of the Galactic FRB 200428 \citep{andersen_bright_2020}.
Nonetheless, upward drifting seems more rare than the opposite while the polar-cap geometry is symmetric in that respect. This could be related to the difficulty or impossibility of measuring drifts in broadband bursts. It could also be due to an asymmetry between the upward drifting and downward drifting regions of the magnetosphere, for example if a toroidal component is present albeit weaker than in repeaters. 

It is thus interesting to consider the possibility that the main difference between repeaters and apparent non-repeaters lies in the strength of their toroidal components. It is noteworthy that the theory of magnetars relies on the dissipation of a strong internal toroidal field into crustal fractures provoking outbursts. Younger objects then presumably have stronger, not yet dissipated, toroidal components which lead to more frequent outbursts \citep{perna_unified_2011}. Very young objects are also expected to spin faster than known magnetars, which corresponds to the case explored in Sec. \ref{sec:dipoletoro} for which a fast spin period of $P=250$ms is favoured to produce an envelope drift rate similar to what is observed from repeaters such as FRB121102 or FRB180916.

Scenarios compatible with the present discussion have been proposed \citep[see e.g.][ and references therein]{zhang_physical_2020}. In these scenarios, repeating sources are more active because they are young fast-spinning magnetars subject to frequent outbursts responsible for triggering FRB emission. On the other hand, non-repeaters form a slower and apparently less active magnetar population, possibly older but not necessarily, with repetitions occurring on a much longer timescale.

Our model cannot currently explain the relation between burst width and fluence noted by several authors. Indeed, repeaters appear to produce longer but less energetic bursts than one-off events \citep[e.g.][]{pleunis_fast_2021}. In our model, these quantities have to be specified ad hoc in the properties of a burst. In a magnetar starquake model, these properties may depend on the starquakes themselves, which is entirely outside of the model. However it might also be related to non-local geometrical aspects, in particular the propagation of the emitting elements from the source to the location visible by the observer. For example, one can speculate that a lower observed fluence and longer duration results from a purely geometrical dilution of the emitting elements in time and directions due to a longer path in the magnetosphere. Computing the global propagation of emitting elements and not only the local one would also in principle permit to relate the effective opening angles at different points of the path with one another. Such calculation is differed to further work. 

Similarly, it is possible that non-repeaters have a low repetition rate not only because they are less active but also because their emission is more collimated than that of repeaters, as has already been proposed by \citet{connor_beaming_2020}. In our framework, this means modelling the effective opening angle. 
In the case of the Galactic magnetar SGR 1935+2154 most X-ray outbursts are not associated with FRB emission \citep{cai_insight-hxmt_2022, chen_frb-srb-xrb_2022} which suggests that either outbursts are not a sufficient condition for FRB emission or that it is highly collimated.

Another obvious extension of the model concerns the use the more realistic or more complex geometries. More realistic geometries include plasma effects, as in force-free magnetospheres \cite[e.g.][]{timokhin_force-free_2006} for instance, while more complex geometries can result from the addition of multipoles or particular magnetospheric currents as in some models of twisted magnetospheres for magnetars \citep{thompson_electrodynamics_2002}. An additional difficulty in the latter case is the transitory nature of these currents \citep{beloborodov_untwisting_2009}. We note here that another, and complementary, way of adding complexity is to extend the formalism to next order in curvilinear abscissa, thus including torsion and curvature variation (see Eq. \eqref{eq:OM}).

There is already evidence of more complex geometries. The occasionally observed polarisation swings from repeater FRB 180301 \citep[e.g.][]{luo_diverse_2020} cannot be reproduced with the geometries studied in this paper. Another interesting case is that of the sub-second periodicity detected in the sub-bursts of FRB20191221A \citep{andersen_sub-second_2022} which, if associated with a spin period, may imply an emission duration much larger than that of a single envelope, and more similar to the continuous emission of pulsars. As discussed in Secs. \ref{sec:bundlereduc} and \ref{sec:propagation}, this situation is not within the scope of this paper as it may not be treated in a local framework. 
Interestingly, much fainter bursts have also been observed from SGR 1935+2154 \citep{kirsten_detection_2020, zhang_highly_2020}. If these follow from the same process as FRBs, which would then make it span 7 orders of magnitude in luminosity, then it must be noted that those cannot come from dipolar polar caps. Indeed, two of these bursts were separated by 1.4s compared to the 3.2s period of the magnetar. This is not either consistent with antipodal polar caps of an orthogonal rotator which would face the observer every half period. This implies that either emission does not occur in a low-altitude open-field region or that the magnetic geometry is more complex than dipolar. The latter option is plausible and has been exemplified by NICER observations of millisecond pulsars \citep{riley_nicer_2021, bilous_nicer_2019}. 

Given the number of parameters involved (see Sec. \ref{sec:applications}), the model can only be quantitatively constraining when fitted to a population of bursts rather than a single event. This strongly favours applications to repeating sources, although it may also be applied to the population of faint bursts from SGR 1935+2154 if one assumes they share the same physical mechanism with FRBs. Thus, as such the two examples developed in Sec. \ref{sec:applications} demonstrate that the model is compatible with single events, but the real test will come from a simultaneous fit to a whole burst population from a single source.

\section{Summary and conclusion \label{sec:conclusion}}
In this paper, we have developed a simple local model where the frequency and polarisation envelopes of FRBs are determined by local geometrical quantities such as the curvature and orientation of a single streamline endowed with an effective emission angle $\Omega$. Geometry was mapped onto observables using the assumption of radius-to-frequency mapping on the one hand, and the rotating vector model for polarisation on the other hand. We call an envelope a closed contour that constrains the burst spectral occupancy as a function of time. Respectively, a polarisation envelope is defined, which delimits the range of the PA values allowed as a function of time. We propose a simple model of the bursts themselves as a function of generic intrinsic quantities such as the injection time and duration, as well as the intrinsic angular profile of the emission process. We, however, do not specify any specific emission process. 

As an example, we have considered the envelopes generated in the dipolar polar-cap geometry in \ref{sec:dipole}, typical of NS magnetospheres, and we have investigated the effects of an added strong counter-rotating toroidal component in Sec. \ref{sec:dipoletoro}. The emitting plasma is assumed to be streaming along magnetic-field lines. We also show examples of bursts filling their envelopes (Figs. \ref{fig:burstJ1935} and \ref{fig:bursttoro}), which we call pseudo-Gaussian because they were generated using a Gaussian intrinsic angular profile as well as a Gaussian injection profile (see Sec. \ref{sec:pseudogaussian}). We stress that the resulting dynamic spectra are not Gaussian and notably display asymmetries, hence the `pseudo'. 

 The variation in the envelope with time yields the drift of the burst characteristic and extremal frequencies. We call it envelope drift in this work. 
 
The spectrum of a single point-like emitting element is a line crossing the envelope from upper to lower frequencies. Finite-speed propagation of emitting elements gives a slope to this spectrum, which we call propagation drift. It is shown to be usually much steeper than the envelope drift and always downward oriented, contrary to the envelope drift. However, we caution that in the model of pseudo-Gaussian bursts, the figure of the burst is asymmetric, which may result in an apparent slope when fitted with empirical models independently of propagation drift (see discussion in Sec. \ref{sec:dipoletoro}). This is due to the burst resulting from a continuum of emitting elements that lasts sufficiently long to be sensitive to the variation of the envelope itself.

In Sec. \ref{sec:dipole}, we have seen that the dipolar polar-cap geometry produces a broad variety of envelopes depending on the location of the emission region with respect to the magnetic pole. We focus on the outer part of the polar cap, where we have shown that this geometry can produce broad-band envelopes with both upward and downward drifts depending on the location of the emitting line with respect to the magnetic pole. On the other hand, the propagation drift was shown to have a negligible effect. In Fig. \ref{fig:burstJ1935}, we show broad-band pseudo-Gaussian bursts truncated into the bandwidth of CHIME/FRB. As a consequence of truncation, no asymmetry of the bursts is noticeable that could be interpreted as a burst `slope'. Importantly, we have shown in Fig. \ref{fig:burstJ1935} that the dipolar geometry can be used to produce bursts qualitatively similar to those observed from the Galactic magnetar SGR J1935+2154  \citep{andersen_bright_2020}, that is FRB200824.

When an additional strong counter-rotating toroidal component is added (Sec. \ref{sec:dipoletoro}), the model generically produces downward banana-shaped envelopes (Fig. \ref{fig:bursttoro}). This means there is narrow-band emission with downward envelope drift, similar to what is commonly observed from repeaters. We have further shown that all else being equal, this geometry requires a faster spin period in order to reproduce envelope drifts of the order of $100 \rm s^{-1}$ commensurate with observations. In this case as well the propagation drift appears too fast to be noticeable by eye, but the narrow-band nature of the envelope makes the asymmetry of bursts quite apparent. This may be measured as a slope when fitted with empirical models. The three bursts shown in Fig. \ref{fig:bursttoro} are meant to illustrate how downward-drifting sub-bursts akin to those of FRB121102 published in \citet{hessels_frb_2019} can be produced by this model. 

In both geometries, we have found that polarisation envelopes are narrow, a few degrees wide at most, and are dominated by a moderate linear drift of the order of $1^{\circ} /\rm ms$.  As a result, the measured PA appears constant within a single event unless an accuracy better than a degree can be achieved. On the other hand, the constant level can vary across a wide range depending on the location in the magnetosphere with which the envelope is associated. Thus, if different events come from different locations, one expects a measurable and possibly large variation in the constant between them. Further, we have shown that for narrow-band bursts, part of the intrinsic polarisation signal can be absorbed into a fit for the RM at a level of a few $\rm rad/m^{2}$ (Eq. \eqref{eq:drm}). Again, this effect is location-dependent, meaning that one expects apparent RM fluctuations between events which may contribute to observed variations from repeating sources. Altogether, polarisation properties appear broadly compatible with observations with the notable exception of the polarisation swing observed from FRB180301 \citep{luo_diverse_2020}.

The cases explored in this paper suggest a relationship between the morphology of FRBs and the relative strength of a toroidal component in the magnetosphere. In addition, the results suggest that a stronger toroidal component should be associated with faster rotation in order to correctly reproduce the observed drift rates. We have briefly discussed how this supports models which associate young magnetars with stronger toroidal fields and faster rotation to repeating FRB sources, and slower and less twisted magnetospheres to apparent non-repeaters. 

We believe that the main interest of our model is to take a step towards a quantitative relationship between burst morphology and the properties of the source. There is still work to be done in order to implement more realistic geometries and construct more physically motivated burst models. Ultimately, the goal is to gain new insight into FRB sources, such as their spin period or magnetospheric properties, by fitting populations of bursts in the time-frequency plane.

\begin{acknowledgements}
        The author thanks T. Francez, Dr. F. Mottez, Dr. I. Cognard, Dr. M. Cruces, and Dr. P. Zarka for helpful and supportive discussions as well as proofreading and comments on the manuscript. 
        
        The author thanks the anonymous referee for their constructive comments. 
        
        The author acknowledges the use of the formal calculation engine Sage, \url{www.sagemath.org}. Plots were made using Python libraries such as Scipy, \url{www.scipy.org}, and Matplotlib, \url{www.matplotlib.org}. 
\end{acknowledgements}

%
\bibliographystyle{aa} 
\bibliography{Geometry22} 

\begin{thebibliography}{69}
\expandafter\ifx\csname natexlab\endcsname\relax\def\natexlab#1{#1}\fi

\bibitem[{Andersen {et~al.}(2019)Andersen, Bandura, Bhardwaj, Boubel, Boyce,
  Boyle, Brar, Cassanelli, Chawla, Cubranic, Deng, Dobbs, Fandino, Fonseca,
  Gaensler, Gilbert, Giri, Good, Halpern, Hill, Hinshaw, H{\"o}fer, Josephy,
  Kaspi, Kothes, Landecker, Lang, Li, Lin, Masui, Mena-Parra, Merryfield,
  Mckinven, Michilli, Milutinovic, Naidu, Newburgh, Ng, Patel, Pen,
  Pinsonneault-Marotte, Pleunis, Rafiei-Ravandi, Rahman, Ransom, Renard,
  Scholz, Siegel, Singh, Smith, Stairs, Tendulkar, Tretyakov, Vanderlinde,
  Yadav, \& Zwaniga}]{andersen_chimefrb_2019}
Andersen, a. B.~C., Bandura, K., Bhardwaj, M., {et~al.} 2019, The Astrophysical
  Journal, 885, L24

\bibitem[{Andersen {et~al.}(2022)Andersen, Bandura, Bhardwaj, Boyle, Brar,
  Breitman, Cassanelli, Chatterjee, Chawla, Cliche, Cubranic, Curtin, Deng,
  Dobbs, Dong, Fonseca, Gaensler, Giri, Good, Hill, Josephy, Kaczmarek, Kader,
  Kania, Kaspi, Leung, Li, Lin, Masui, Mckinven, Mena-Parra, Merryfield,
  Meyers, Michilli, Naidu, Newburgh, Ng, Ordog, Patel, Pearlman, Pen, Petroff,
  Pleunis, Rafiei-Ravandi, Rahman, Ransom, Renard, Sanghavi, Scholz, Shaw,
  Shin, Siegel, Singh, Smith, Stairs, Tan, Tendulkar, Vanderlinde, Wiebe, Wulf,
  \& Zwaniga}]{andersen_sub-second_2022}
Andersen, B.~C., Bandura, K., Bhardwaj, M., {et~al.} 2022, Nature, 607, 256,
  number: 7918 Publisher: Nature Publishing Group

\bibitem[{Andersen {et~al.}(2020)Andersen, Bandura, Bhardwaj, Bij, Boyce,
  Boyle, Brar, Cassanelli, Chawla, Chen, Cliche, Cook, Cubranic, Curtin,
  Denman, Dobbs, Dong, Fandino, Fonseca, Gaensler, Giri, Good, Halpern, Hill,
  Hinshaw, H{\"o}fer, Josephy, Kania, Kaspi, Landecker, Leung, Li, Lin, Masui,
  Mckinven, Mena-Parra, Merryfield, Meyers, Michilli, Milutinovic, Mirhosseini,
  M{\"u}nchmeyer, Naidu, Newburgh, Ng, Patel, Pen, Pinsonneault-Marotte,
  Pleunis, Quine, Rafiei-Ravandi, Rahman, Ransom, Renard, Sanghavi, Scholz,
  Shaw, Shin, Siegel, Singh, Smegal, Smith, Stairs, Tan, Tendulkar, Tretyakov,
  Vanderlinde, Wang, Wulf, Zwaniga, \& {The CHIME/FRB
  Collaboration}}]{andersen_bright_2020}
Andersen, B.~C., Bandura, K.~M., Bhardwaj, M., {et~al.} 2020, Nature, 587, 54,
  number: 7832 Publisher: Nature Publishing Group

\bibitem[{Anna-Thomas {et~al.}(2022)Anna-Thomas, Connor, Burke-Spolaor,
  Beniamini, Aggarwal, Law, Lynch, Li, Feng, Ocker, Cruces, Chatterjee, Yu,
  Niu, \& Xue}]{anna-thomas_highly_2022}
Anna-Thomas, R., Connor, L., Burke-Spolaor, S., {et~al.} 2022, A {Highly}
  {Variable} {Magnetized} {Environment} in a {Fast} {Radio} {Burst} {Source},
  arXiv:2202.11112 [astro-ph]

\bibitem[{Bagchi(2017)}]{bagchi_unified_2017}
Bagchi, M. 2017, The Astrophysical Journal, 838, L16, publisher: American
  Astronomical Society

\bibitem[{Barr{\`e}re {et~al.}(2022)Barr{\`e}re, Guilet, Reboul-Salze, Raynaud,
  \& Janka}]{barrere_new_2022}
Barr{\`e}re, P., Guilet, J., Reboul-Salze, A., Raynaud, R., \& Janka, H.-T.
  2022, Astronomy and Astrophysics, 668, A79

\bibitem[{Beloborodov(2009)}]{beloborodov_untwisting_2009}
Beloborodov, A.~M. 2009, The Astrophysical Journal, 703, 1044

\bibitem[{Beloborodov(2017)}]{beloborodov_flaring_2017}
Beloborodov, A.~M. 2017, The Astrophysical Journal Letters, 843, L26,
  publisher: The American Astronomical Society

\bibitem[{Beloborodov(2020)}]{beloborodov_blast_2020}
Beloborodov, A.~M. 2020, The Astrophysical Journal, 896, 142, publisher: The
  American Astronomical Society

\bibitem[{Bilous {et~al.}(2019)Bilous, Watts, Harding, Riley, Arzoumanian,
  Bogdanov, Gendreau, Ray, Guillot, Ho, \& Chakrabarty}]{bilous_nicer_2019}
Bilous, A.~V., Watts, A.~L., Harding, A.~K., {et~al.} 2019, The Astrophysical
  Journal Letters, 887, L23

\bibitem[{Bochenek {et~al.}(2020)Bochenek, Ravi, Belov, Hallinan, Kocz,
  Kulkarni, \& McKenna}]{bochenek_fast_2020}
Bochenek, C.~D., Ravi, V., Belov, K.~V., {et~al.} 2020, Nature, 587, 59

\bibitem[{Cai {et~al.}(2022)Cai, Xue, Li, Xiong, Zhang, Lin, Li, Ge, Zhao,
  Song, Lu, Zhang, Zhang, Xiao, Tuo, Yi, Guo, Xie, Zhao, Zhang, Li, Liu, Zheng,
  \& Wang}]{cai_insight-hxmt_2022}
Cai, C., Xue, W.-C., Li, C.-K., {et~al.} 2022, The Astrophysical Journal
  Supplement Series, 260, 24, publisher: The American Astronomical Society

\bibitem[{Caleb \& Keane(2021)}]{caleb_decade_2021}
Caleb, M. \& Keane, E. 2021, Universe, 7, 453, number: 11 Publisher:
  Multidisciplinary Digital Publishing Institute

\bibitem[{Chamma {et~al.}(2022)Chamma, Rajabi, Kumar, \&
  Houde}]{chamma_broad_2022}
Chamma, M.~A., Rajabi, F., Kumar, A., \& Houde, M. 2022, A broad survey of
  spectro-temporal properties from {FRB20121102A}, arXiv:2210.00106 [astro-ph]

\bibitem[{Chen \& Zhang(2022)}]{chen_frb-srb-xrb_2022}
Chen, C.~J. \& Zhang, B. 2022, {FRB}\$-\${SRB}\$-\${XRB}: {Geometric} and
  {Relativistic} {Beaming} {Constraints} of {Fast} {Radio} {Bursts} from the
  {Galactic} {Magnetar} {SGR} {J1935}+2154, arXiv:2210.01904 [astro-ph]

\bibitem[{Collaboration {et~al.}(2020)Collaboration, Amiri, Andersen, Bandura,
  Bhardwaj, Boyle, Brar, Chawla, Chen, Cliche, Cubranic, Deng, Denman, Dobbs,
  Dong, Fandino, Fonseca, Gaensler, Giri, Good, Halpern, Hessels, Hill,
  H{\"o}fer, Josephy, Kania, Karuppusamy, Kaspi, Keimpema, Kirsten, Landecker,
  Lang, Leung, Li, Lin, Marcote, Masui, Mckinven, Mena-Parra, Merryfield,
  Michilli, Milutinovic, Mirhosseini, Naidu, Newburgh, Ng, Nimmo, Paragi,
  Patel, Pen, Pinsonneault-Marotte, Pleunis, Rafiei-Ravandi, Rahman, Ransom,
  Renard, Sanghavi, Scholz, Shaw, Shin, Siegel, Singh, Smegal, Smith, Stairs,
  Tendulkar, Tretyakov, Vanderlinde, Wang, Wang, Wulf, Yadav, \&
  Zwaniga}]{collaboration_periodic_2020}
Collaboration, T.~C., Amiri, M., Andersen, B.~C., {et~al.} 2020,
  arXiv:2001.10275 [astro-ph], arXiv: 2001.10275

\bibitem[{Connor {et~al.}(2020)Connor, Miller, \&
  Gardenier}]{connor_beaming_2020}
Connor, L., Miller, M.~C., \& Gardenier, D.~W. 2020, {\textbackslash}mnras,
  497, 3076, \_eprint: 2003.11930

\bibitem[{Cordes(1978)}]{cordes_observational_1978}
Cordes, J.~M. 1978, The Astrophysical Journal, 222, 1006

\bibitem[{Cordes \& Wasserman(2016)}]{cordes_supergiant_2016}
Cordes, J.~M. \& Wasserman, I. 2016, {\textbackslash}mnras, 457, 232, \_eprint:
  1501.00753

\bibitem[{Cruces {et~al.}(2021)Cruces, Spitler, Scholz, Lynch, Seymour,
  Hessels, Gouiff{\'e}s, Hilmarsson, Kramer, \& Munjal}]{cruces_repeating_2021}
Cruces, M., Spitler, L.~G., Scholz, P., {et~al.} 2021, Monthly Notices of the
  Royal Astronomical Society, 500, 448

\bibitem[{Dai(2020)}]{dai_magnetar-asteroid_2020}
Dai, Z.~G. 2020, {\textbackslash}apjl, 897, L40, \_eprint: 2005.12048

\bibitem[{Dai {et~al.}(2016)Dai, Wang, Wu, \& Huang}]{dai_repeating_2016}
Dai, Z.~G., Wang, J.~S., Wu, X.~F., \& Huang, Y.~F. 2016, {\textbackslash}apj,
  829, 27, \_eprint: 1603.08207

\bibitem[{Dai \& Zhong(2020)}]{dai_periodic_2020}
Dai, Z.~G. \& Zhong, S.~Q. 2020, {\textbackslash}apjl, 895, L1, \_eprint:
  2003.04644

\bibitem[{Deutsch(1955)}]{deutsch_electromagnetic_1955}
Deutsch, A.~J. 1955, Annales d'Astrophysique, 18, 1

\bibitem[{Farah {et~al.}(2018)Farah, Flynn, Bailes, Jameson, Bannister, Barr,
  Bateman, Bhandari, Caleb, Campbell-Wilson, Chang, Deller, Green, Hunstead,
  Jankowski, Keane, Macquart, M{\"o}ller, Onken, Os{\l }owski, Parthasarathy,
  Plant, Ravi, Shannon, Tucker, Venkatraman~Krishnan, \& Wolf}]{farah_frb_2018}
Farah, W., Flynn, C., Bailes, M., {et~al.} 2018, Monthly Notices of the Royal
  Astronomical Society, 478, 1209

\bibitem[{Goldreich \& Julian(1969)}]{goldreich_pulsar_1969}
Goldreich, P. \& Julian, W.~H. 1969, The Astrophysical Journal, 157, 869

\bibitem[{Hessels {et~al.}(2019)Hessels, Spitler, Seymour, Cordes, Michilli,
  Lynch, Gourdji, Archibald, Bassa, Bower, Chatterjee, Connor, Crawford,
  Deneva, Gajjar, Kaspi, Keimpema, Law, Marcote, McLaughlin, Paragi, Petroff,
  Ransom, Scholz, Stappers, \& Tendulkar}]{hessels_frb_2019}
Hessels, J. W.~T., Spitler, L.~G., Seymour, A.~D., {et~al.} 2019, The
  Astrophysical Journal, 876, L23, publisher: American Astronomical Society

\bibitem[{Hilmarsson {et~al.}(2021)Hilmarsson, Spitler, Main, \&
  Li}]{hilmarsson_polarization_2021}
Hilmarsson, G.~H., Spitler, L.~G., Main, R.~A., \& Li, D.~Z. 2021, Monthly
  Notices of the Royal Astronomical Society, 508, 5354

\bibitem[{Jahns {et~al.}(2022)Jahns, Spitler, Nimmo, Hewitt, Snelders, Seymour,
  Hessels, Gourdji, Michilli, \& Hilmarsson}]{jahns_frb_2022}
Jahns, J.~N., Spitler, L.~G., Nimmo, K., {et~al.} 2022, The {FRB} {20121102A}
  {November} rain in 2018 observed with the {Arecibo} {Telescope},
  arXiv:2202.05705 [astro-ph]

\bibitem[{Josephy {et~al.}(2019)Josephy, Chawla, Fonseca, Ng, Patel, Pleunis,
  Scholz, Andersen, Bandura, Bhardwaj, Boyce, Boyle, Brar, Cubranic, Dobbs,
  Gaensler, Gill, Giri, Good, Halpern, Hinshaw, Kaspi, Landecker, Lang, Lin,
  Masui, Mckinven, Mena-Parra, Merryfield, Michilli, Milutinovic, Naidu, Pen,
  Rafiei-Ravandi, Rahman, Ransom, Renard, Siegel, Smith, Stairs, Tendulkar,
  Vanderlinde, Yadav, \& Zwaniga}]{2019ApJ...882L..18J}
Josephy, A., Chawla, P., Fonseca, E., {et~al.} 2019, The Astrophysical Journal,
  882, L18, aDS Bibcode: 2019ApJ...882L..18J

\bibitem[{Kirsten {et~al.}(2020)Kirsten, Snelders, Jenkins, Nimmo, van~den
  Eijnden, Hessels, Gawro{\'n}ski, \& Yang}]{kirsten_detection_2020}
Kirsten, F., Snelders, M.~P., Jenkins, M., {et~al.} 2020, Nature Astronomy, 1,
  publisher: Nature Publishing Group

\bibitem[{Komesaroff(1970)}]{komesaroff_possible_1970}
Komesaroff, M.~M. 1970, Nature, 225, 612

\bibitem[{Kumar {et~al.}(2017)Kumar, Lu, \& Bhattacharya}]{kumar_fast_2017}
Kumar, P., Lu, W., \& Bhattacharya, M. 2017, Monthly Notices of the Royal
  Astronomical Society, 468, 2726

\bibitem[{Lanman {et~al.}(2022)Lanman, Andersen, Chawla, Josephy, Noble, Kaspi,
  Bandura, Bhardwaj, Boyle, Brar, Breitman, Cassanelli, Dong, Fonseca,
  Gaensler, Good, Kaczmarek, Leung, Masui, Meyers, Ng, Patel, Pearlman,
  Petroff, Pleunis, Rafiei-Ravandi, Rahman, Sanghavi, Scholz, Shin, Stairs,
  Tendulkar, \& Zwaniga}]{lanman_sudden_2022}
Lanman, A.~E., Andersen, B.~C., Chawla, P., {et~al.} 2022, The Astrophysical
  Journal, 927, 59, publisher: The American Astronomical Society

\bibitem[{Lu {et~al.}(2020)Lu, Kumar, \& Zhang}]{lu_unified_2020}
Lu, W., Kumar, P., \& Zhang, B. 2020, {\textbackslash}mnras, 498, 1397,
  \_eprint: 2005.06736

\bibitem[{Luo {et~al.}(2020)Luo, Wang, Men, Zhang, Jiang, Xu, Wang, Lee, Han,
  Zhang, Caballero, Chen, Chen, Gan, Guo, Hao, Huang, Jiang, Li, Li, Li, Luo,
  Pan, Pei, Qian, Sun, Wang, Wang, Wen, Xu, Xu, Yan, Yan, Yu, Yuan, Zhang, \&
  Zhu}]{luo_diverse_2020}
Luo, R., Wang, B.~J., Men, Y.~P., {et~al.} 2020, Nature, 586, 693

\bibitem[{Lyubarsky(2014)}]{lyubarsky_model_2014}
Lyubarsky, Y. 2014, Monthly Notices of the Royal Astronomical Society, 442, L9

\bibitem[{Lyubarsky(2020)}]{lyubarsky_fast_2020}
Lyubarsky, Y. 2020, The Astrophysical Journal, 897, 1, publisher: American
  Astronomical Society

\bibitem[{Lyutikov(2020)}]{lyutikov_radius--frequency_2020}
Lyutikov, M. 2020, {\textbackslash}apj, 889, 135, \_eprint: 1909.10409

\bibitem[{Majid {et~al.}(2021)Majid, Pearlman, Prince, Wharton, Naudet, Bansal,
  Connor, Bhardwaj, \& Tendulkar}]{majid_bright_2021}
Majid, W.~A., Pearlman, A.~B., Prince, T.~A., {et~al.} 2021, The Astrophysical
  Journal Letters, 919, L6

\bibitem[{Makishima {et~al.}(2021)Makishima, Tamba, Aizawa, Odaka, Yoneda,
  Enoto, \& Suzuki}]{makishima_discovery_2021}
Makishima, K., Tamba, T., Aizawa, Y., {et~al.} 2021, The Astrophysical Journal,
  923, 63, aDS Bibcode: 2021ApJ...923...63M

\bibitem[{Mckinven {et~al.}(2022)Mckinven, Gaensler, Michilli, Masui, Kaspi,
  Bhardwaj, Cassanelli, Chawla, Dong, Fonseca, Leung, Li, Ng, Patel, Petroff,
  Pearlman, Pleunis, Rafiei-Ravandi, Rahman, Sand, Shin, Scholz, Stairs, Smith,
  Su, \& Tendulkar}]{mckinven_large_2022}
Mckinven, R., Gaensler, B.~M., Michilli, D., {et~al.} 2022, A {Large} {Scale}
  {Magneto}-ionic {Fluctuation} in the {Local} {Environment} of {Periodic}
  {Fast} {Radio} {Burst} {Source}, {FRB} {20180916B}, arXiv:2205.09221
  [astro-ph]

\bibitem[{Metzger {et~al.}(2017)Metzger, Berger, \&
  Margalit}]{metzger_millisecond_2017}
Metzger, B.~D., Berger, E., \& Margalit, B. 2017, The Astrophysical Journal,
  841, 14

\bibitem[{Michel(1973)}]{michel_rotating_1973}
Michel, F.~C. 1973, The Astrophysical Journal, 180, 207

\bibitem[{Michilli {et~al.}(2018)Michilli, Seymour, Hessels, Spitler, Gajjar,
  Archibald, Bower, Chatterjee, Cordes, Gourdji, Heald, Kaspi, Law, Sobey,
  Adams, Bassa, Bogdanov, Brinkman, Demorest, Fernandez, Hellbourg, Lazio,
  Lynch, Maddox, Marcote, McLaughlin, Paragi, Ransom, Scholz, Siemion,
  Tendulkar, Rooy, Wharton, \& Whitlow}]{michilli_extreme_2018}
Michilli, D., Seymour, A., Hessels, J. W.~T., {et~al.} 2018, Nature, 553, 182

\bibitem[{Mottez \& Zarka(2014)}]{mottez_radio_2014}
Mottez, F. \& Zarka, P. 2014, Astronomy and Astrophysics, 569, A86

\bibitem[{Mottez {et~al.}(2020)Mottez, Zarka, \&
  Voisin}]{mottez_repeating_2020}
Mottez, F., Zarka, P., \& Voisin, G. 2020, Astronomy \& Astrophysics, 644,
  A145, publisher: EDP Sciences

\bibitem[{Nimmo {et~al.}(2021)Nimmo, Hessels, Keimpema, Archibald, Cordes,
  Karuppusamy, Kirsten, Li, Marcote, \& Paragi}]{nimmo_highly_2021}
Nimmo, K., Hessels, J. W.~T., Keimpema, A., {et~al.} 2021, Nature Astronomy, 5,
  594, aDS Bibcode: 2021NatAs...5..594N

\bibitem[{Perna \& Pons(2011)}]{perna_unified_2011}
Perna, R. \& Pons, J.~A. 2011, The Astrophysical Journal Letters, 727, L51

\bibitem[{P{\'e}tri(2016)}]{petri_theory_2016}
P{\'e}tri, J. 2016, Journal of Plasma Physics, 82, 635820502

\bibitem[{P{\'e}tri(2017)}]{petri_polarized_2017}
P{\'e}tri, J. 2017, Monthly Notices of the Royal Astronomical Society, 466, L73

\bibitem[{Petroff {et~al.}(2022)Petroff, Hessels, \&
  Lorimer}]{petroff_fast_2022}
Petroff, E., Hessels, J. W.~T., \& Lorimer, D.~R. 2022, Astronomy and
  Astrophysics Review, 30, 2

\bibitem[{Pleunis {et~al.}(2021)Pleunis, Good, Kaspi, Mckinven, Ransom, Scholz,
  Bandura, Bhardwaj, Boyle, Brar, Cassanelli, Chawla, (Adam)~Dong, Fonseca,
  Gaensler, Josephy, Kaczmarek, Leung, Lin, Masui, Mena-Parra, Michilli, Ng,
  Patel, Rafiei-Ravandi, Rahman, Sanghavi, Shin, Smith, Stairs, \&
  Tendulkar}]{pleunis_fast_2021}
Pleunis, Z., Good, D.~C., Kaspi, V.~M., {et~al.} 2021, The Astrophysical
  Journal, 923, 1, aDS Bibcode: 2021ApJ...923....1P

\bibitem[{Popov \& Postnov(2013)}]{popov_millisecond_2013}
Popov, S.~B. \& Postnov, K.~A. 2013, arXiv e-prints, 1307, arXiv:1307.4924

\bibitem[{Radhakrishnan \& Cooke(1969)}]{radhakrishnan_magnetic_1969}
Radhakrishnan, V. \& Cooke, D.~J. 1969, Astrophysical Letters, 3, 225

\bibitem[{Rajwade {et~al.}(2020)Rajwade, Mickaliger, Stappers, Morello,
  Agarwal, Bassa, Breton, Caleb, Karastergiou, Keane, \&
  Lorimer}]{rajwade_possible_2020}
Rajwade, K.~M., Mickaliger, M.~B., Stappers, B.~W., {et~al.} 2020, Monthly
  Notices of the Royal Astronomical Society, 495, 3551

\bibitem[{Riley {et~al.}(2021)Riley, Watts, Ray, Bogdanov, Guillot, Morsink,
  Bilous, Arzoumanian, Choudhury, Deneva, Gendreau, Harding, Ho, Lattimer,
  Loewenstein, Ludlam, Markwardt, Okajima, Prescod-Weinstein, Remillard, Wolff,
  Fonseca, Cromartie, Kerr, Pennucci, Parthasarathy, Ransom, Stairs, Guillemot,
  \& Cognard}]{riley_nicer_2021}
Riley, T.~E., Watts, A.~L., Ray, P.~S., {et~al.} 2021, The Astrophysical
  Journal, 918, L27, aDS Bibcode: 2021ApJ...918L..27R

\bibitem[{Rodr{\'i}guez~Castillo {et~al.}(2016)Rodr{\'i}guez~Castillo, Israel,
  Tiengo, Salvetti, Turolla, Zane, Rea, Esposito, Mereghetti, Perna, Stella,
  Pons, Campana, G{\"o}tz, \& Motta}]{rodriguez_castillo_outburst_2016}
Rodr{\'i}guez~Castillo, G.~A., Israel, G.~L., Tiengo, A., {et~al.} 2016,
  Monthly Notices of the Royal Astronomical Society, 456, 4145

\bibitem[{Smallwood {et~al.}(2019)Smallwood, Martin, \&
  Zhang}]{smallwood_investigation_2019}
Smallwood, J.~L., Martin, R.~G., \& Zhang, B. 2019, {\textbackslash}mnras, 485,
  1367, \_eprint: 1902.05203

\bibitem[{Suresh {et~al.}(2021)Suresh, Cordes, Chatterjee, Gajjar, Perez,
  Siemion, \& Price}]{suresh_48_2021}
Suresh, A., Cordes, J.~M., Chatterjee, S., {et~al.} 2021, The Astrophysical
  Journal, 921, 101

\bibitem[{Thompson \& Duncan(1996)}]{thompson_soft_1996}
Thompson, C. \& Duncan, R.~C. 1996, The Astrophysical Journal, 473, 322

\bibitem[{Thompson {et~al.}(2002)Thompson, Lyutikov, \&
  Kulkarni}]{thompson_electrodynamics_2002}
Thompson, C., Lyutikov, M., \& Kulkarni, S.~R. 2002, The Astrophysical Journal,
  574, 332, publisher: IOP Publishing

\bibitem[{Timokhin(2006)}]{timokhin_force-free_2006}
Timokhin, A.~N. 2006, Monthly Notices of the Royal Astronomical Society, 368,
  1055

\bibitem[{Voisin {et~al.}(2021)Voisin, Mottez, \& Zarka}]{voisin_periodic_2021}
Voisin, G., Mottez, F., \& Zarka, P. 2021, Monthly Notices of the Royal
  Astronomical Society, 508, 2079

\bibitem[{Wadiasingh \& Chirenti(2020)}]{wadiasingh_fast_2020}
Wadiasingh, Z. \& Chirenti, C. 2020, {\textbackslash}apjl, 903, L38, \_eprint:
  2006.16231

\bibitem[{Wadiasingh \& Timokhin(2019)}]{wadiasingh_repeating_2019}
Wadiasingh, Z. \& Timokhin, A. 2019, {\textbackslash}apj, 879, 4, \_eprint:
  1904.12036

\bibitem[{Yuan {et~al.}(2020)Yuan, Beloborodov, Chen, \&
  Levin}]{yuan_plasmoid_2020}
Yuan, Y., Beloborodov, A.~M., Chen, A.~Y., \& Levin, Y. 2020, The Astrophysical
  Journal, 900, L21, publisher: American Astronomical Society

\bibitem[{Zhang(2020)}]{zhang_physical_2020}
Zhang, B. 2020, Nature, 587, 45, number: 7832 Publisher: Nature Publishing
  Group

\bibitem[{Zhang {et~al.}(2020)Zhang, Jiang, Men, Wang, Xu, Xu, Niu, Zhou, Guan,
  Han, Jiang, Lee, Li, Lin, Niu, Wang, Wang, Xu, Yu, Zhang, \&
  Zhu}]{zhang_highly_2020}
Zhang, C.~F., Jiang, J.~C., Men, Y.~P., {et~al.} 2020, The Astronomer's
  Telegram, 13699, 1, aDS Bibcode: 2020ATel13699....1Z

\end{thebibliography}
%

%
\begin{appendix}
\section{Basic properties of dipolar field line near the magnetic pole \label{ap:dipolegeom}}
At low altitude, it is usually assumed \cite[e.g.][]{goldreich_pulsar_1969} that the magnetic field is mainly sourced by currents within the star, rather than by magnetospheric plasma. In addition, the co-rotation velocity is negligible compared to the speed of light, or equivalently $r\sin\theta \ll R_L$, and retardations effects due to rotation \citep{deutsch_electromagnetic_1955} can be neglected as well. Under these conditions, a vacuum dipolar magnetic field can be described by the static dipole formula in the rotating frame of the star, 
\begin{equation}
\vB= \frac{B_L}{x^3} \left(3\vx/x -\vn\right).
\end{equation}
In this form, $ x = r/R_L$ and $B_L$ gives the scale of the magnetic field at the light cylinder, and $\vn$ is a unit vector giving the direction of the magnetic moment. 

The magnetic field is symmetric with respect to rotations around the magnetic axis. Thus, field lines can be expressed in a meridian plane as a function of the colatitude and radius. We consider a field line passing through the location at $(x_0, \alpha_0)$, respectively its (normalised) radius and colatitude. Then one can show that the field is described by the parametric relation
\begin{equation}
        x = \frac{x_0}{\sin^2\alpha_0 }\sin^2\alpha.
\end{equation}
The maximum distance reached by the line occurs at $\alpha = \pi/2$ and
\begin{equation}
\label{apeq:xmax}
        x_{\max} =  \frac{x_0}{\sin^2\alpha_0 }.
\end{equation}

In a rotating dipole, field lines crossing the light-cylinder form a bundle of open field lines. The feet of these lines at the surface of the star are describe the polar cap. The exact shape of this bundle depends on magnetospheric currents and retardations effects, but an approximation can be easily obtained by considering the lines that reach a distance $x_{\max}\geq 1$ in the static dipole approximation. Solving for $x_{\max} =1$ using Eq. \eqref{apeq:xmax}, the limit of the bundle at an altitude $x$ is at magnetic colatitude
\begin{equation}
        \label{apeq:alphao}
        \alpha_{\rm o} = \sqrt{x},
\end{equation}
where it is assumed that $\alpha_{\rm o} \ll 1$. We note that at the surface of the star $x=R_*/R_L$ and we recover the usual estimate of the polar cap opening angle of an aligned dipole.

\section{Envelope with instantaneous propagation}\label{ap:instenv}
Inserting Eq. \eqref{eq:zetapm} into Eq.\eqref{eq:r} and keeping only leading order terms in $\zeta$ we obtain the boundaries of the instantaneously visible segment along the radial direction at a given instant $t$,
\begin{equation}
\label{eq:rpm}
r_\pm  = r_c \pm \frac{\Delta r}{2},
\end{equation}
where $r_c = \zeta_c \crho$, $\Delta r = \Delta\zeta \crho$ and $\crho \equiv \ve_r' \cdot \ve_1|_0$. Using Eq. \eqref{eq:zetapm2}, we obtain the first order expressions
\begin{eqnarray}
\label{eq:rc1}
r_c & =& r -  \epsilon h\tmu\bphi, \\
\label{eq:dr1}
\Delta r & = & 2h  \sqrt{1-\bphi^2},
\end{eqnarray}
where $h =  \kappa^{-1} \Omega \crho$ is the characteristic heigh scale of the bundle and, as in Eq.\eqref{eq:zetapm2}, $\bphi \in [-1,1]$ while the bundle is visible. 

Following Eq. \eqref{eq:fpm}, we define the instantaneous visibility envelope at $\vr$ as the region $\{\bar\phi \in [-1,1]; f\in[f_-(\bar\phi), f_+(\bar\phi)]\}_{\vr}$ where 
\begin{eqnarray}
\label{apeq:fpm}
f_\pm(\bar\phi) & = & f\left(\max\left(r_\mp, R_*\right)\right), 
\end{eqnarray}
where $R_*$ is the radius of the star, $r_\pm$ is taken from Eq. \eqref{eq:rpm}, and $f()$ is defined in Eq. \eqref{eq:fr}. In practice we use the fiducial value $R_* = 12$ km. This definition incorporates the fact that the emission region cannot extend below the surface of the star, which translates into an upper frequency cut-off. Since the absolute scaling is not known, in practice we consider the normalised quantity $f_\pm/f_c|_0$ where $f_c|_0=f_c(\bar\phi=0)$. Similarly, we define the central frequency of the envelope by 
\begin{equation}
\label{eq:fc}
f_c(\bar\phi)  =  f\left(\max\left(r_c, R_*\right)\right). 
\end{equation}

Using Eq. \eqref{eq:fr}, we can define the relative spectral occupancy $\Delta f/f_c \equiv \left|f(r_+) - f(r_-)\right| / f(r_c)$. In the case $\Delta r \ll r_c$ and $r_c > R_*$ (or equivalently $\Delta f \ll f_c$), it reduces to  
\begin{equation}
\label{eq:dffc}
\frac{\Delta f}{f_c} = \beta \frac{ 2h  \sqrt{1-\bphi^2}}{r - \epsilon h\tmu \bphi},
\end{equation}
where we used Eqs. \eqref{eq:rc1}-\eqref{eq:dr1}. We note that for the condition $\Delta r \ll r_c$ to be valid implies that $h \ll r$.

Similarly, we define the relative frequency drifting by $\dot f_c/f_c$, 
\begin{equation}
\label{eq:fcdotfc}
\frac{\dot f_c}{f_c} = \beta \omega \smu \frac{h\schi/\Omega}{ r -\epsilon h\tmu \bphi}
,\end{equation}
where we used  $\dot \bphi = \omega |\cmu \schi|/\Omega$ and again the first order expressions  \eqref{eq:rc1}-\eqref{eq:dr1}.

\section{Application to dipolar polar-cap geometry} \label{apsec:dipole}
In the limit of negligible magnetic colatitude, $\alpha \lesssim \alpha_{\rm o} \ll 1$, we normalise the dipolar magnetic field, Eq. \eqref{eq:Bdip}, to obtain the vector field
\begin{equation}
\label{apeq:b}
\vb = \frac{3}{2} \ve_r' - \frac{\vn}{2}.
\end{equation}
As said in the main text and without loss of generality we take $\vn$ in the plane $(\ve_x', \ve_z')$. 

At $\vr$ we have the local Serret-Frenet triad,
\begin{eqnarray}
\label{apeq:e1}
\ve_1|_0 & =& \vb(\vr), \\
\label{apeq:e2}
\kappa \ve_2|_0 & = & \vb\cdot\vec{\nabla}\vb(\vr),\\
\label{apeq:e3}
\ve_3|_0 & =& \ve_1|_0 \times \ve_2|_0.
\end{eqnarray} 
The first vector of the triad is the tangent vector of Eq. \eqref{apeq:b}. 
The second vector of the triad is given by the variation of the tangent vector along itself, $\kappa\ve_2 = \vb\cdot\nabla\vb$, where $\kappa$ is interpreted as the curvature of the line.

At altitudes low compared to the light-cylinder radius it is clear that $\alpha \ll 1$. Magnetic colatitude can be related to spherical coordinates $(\theta,\varphi)$ in $F'$ using $\cos\alpha = \ve_r'\cdot \vn$, where as before $\theta$ is the colatitude taken with respect to $\ve_z'$ . Taylor expanding $\ve_r'$ around $\vn$, that is to say for $\theta = i + \delta\theta, \varphi = 0 + \delta\varphi$, we obtain at leading order
\begin{equation}
\label{apeq:alpha}
\alpha^2 = \delta\theta^2 + \sin^2 i \delta\varphi^2.
\end{equation}

Curvature is computed as the norm of Eq. \eqref{apeq:e2}, the projection of the local tangent onto the radial direction, $\cos\rho$ in Eq. \eqref{eq:rc1} is straightforwardly obtained from \eqref{apeq:e1}, and the height scale $h$ of  Eq. \eqref{eq:rc1} derives from the previous two. Inserting Eqs. \eqref{apeq:e1}-\eqref{apeq:e3} into Eqs. \eqref{eq:chi}-\eqref{eq:sinmu} one further gets the inclination angle $\chi$ and the curvature-plane angle $\mu$. Altogether and at leading order in $\delta\theta \sim \delta\varphi\sim\alpha$  one derives
\begin{eqnarray}
\label{apeq:kappadipole}
\kappa & = & \frac{3}{4R_L \sqrt{x x_{\max}}}, \\
\cos\rho & = & 1, \\
\label{apeq:hdipole}
h & = & \frac{4}{3} R_L  \alpha \Omega x_{\max}, \\
\chi & = & i + \frac{3}{2}\delta\theta, \\
\label{apeq:dipcosmu}
\cos\mu & = & \frac{\delta\theta}{\alpha}, \\
\label{apeq:dipsinmu}
\sin\mu & = & \sin i\frac{\delta\varphi}{\alpha}, \\
\label{apeq:tanmudipole}
\tan\mu & = & \sin i \frac{\delta \varphi}{\delta \theta}, 
\end{eqnarray}
where $x_{\max} = x/\alpha^2$ is the largest distance reached by the field line passing at 
$\vr$ as shown in Eq. \eqref{apeq:xmax}. 

We now show that propagation can be neglected in a large part of the region above the polar cap. One can write the reduced curvature $\bar\kappa = R_L\kappa$ as 
\begin{equation}
\bar\kappa \simeq 63\frac{\bar\alpha}{\sqrt{\bar r}} \left(\frac{P}{1\rm s}\right)^{1/2},
\end{equation}
where $\bar\alpha = \alpha /\alpha_{\rm o}(x)$ gives the colatitude in units of the local polar-cap radius, and  $\bar r = r/R_*$ gives the altitude in units of stellar radius. Thus, as long as the spin period is $P\gg 1$ ms,  magnetic colatitude is not too close to the centre of the polar cap, and altitude is not higher that a few stellar radii, one has $\bar\kappa \gg 1$. 
Using Eq. \eqref{eq:propagphipm}, we see that in the limit $\bar\kappa \rightarrow \infty$, $\Delta t_p = \Delta t \sin\chi|\cos\mu|/\bar\kappa$ such that $\Delta t_p \ll \Delta t$ which is the limit of instantaneous propagation. In this limit one further derives from Eq. \eqref{eq:Deltata} and \eqref{eq:tipm} that $\Delta t_a = \Delta t_i = \Delta t$.

\section{Condition of visibility of an emitting element}\label{ap:propagvis}
We define the pair of functions 
\begin{equation}
        g_{\pm} (\bphi) = a \bphi  - b t_i \mp \sqrt{1 - \bphi^2},
\end{equation}
where $a = \epsilon(\bkappa /(\sin\chi\cos\mu) + \tan\mu)$ and $b = \omega \bkappa /\Omega$ as defined at Eq. \eqref{eq:propagphipm}.

The condition of visibility of an element, Eq. \eqref{eq:propagvis}, can be rewritten as
\begin{equation}
\label{eqap:pvis2}
        g_+(\bphi) \leq 0 \text{ and } g_-(\bphi) \geq 0, 
\end{equation}
which corresponds to $\zeta_p(t,t_i) \leq \zeta_+(t)$ and $\zeta_-(t) \leq \zeta(t,t_i)$, respectively.

It is clear that solutions of \eqref{eqap:pvis2} are also solution of 
\begin{equation}
\label{eqap:gg}
        g_-(\bphi)g_+(\bphi) \leq 0.
\end{equation}
Now, the other solution of Eq. \eqref{eqap:gg} is $\{\bphi \,| \, g_+(\bphi) > 0 \text{ and } g_-(\bphi) < 0\}$. However, we see that by construction $g_-(\bphi) \geq g_+(\bphi)$. Therefore this set is empty and Eq. \eqref{eqap:pvis2} is equivalent to \eqref{eqap:gg}. 

Thanks to this property the problem reduces to the roots of a second-degree polynomial. Indeed, 
\begin{equation}
\label{eqap:gg2}
        g_-(\bphi)g_+(\bphi) = (1+a^2) \bphi^2 - 2ab t_i \bphi + b^2t_i^2 - 1.
\end{equation}
Since the highest-order coefficient is $1+a^2 > 0$, the solution of Eq. \eqref{eqap:gg} is given by $\bphi \in[\bphi_-, \bphi_+]$ where $\bphi_\pm$ are the roots of \eqref{eqap:gg2},
\begin{equation}
        \bphi_\pm = \frac{abt_i \pm \sqrt{1+a^2 - b^2t_i^2}}{1+a^2}.
\end{equation}
The solution can be expressed in terms of times, $t_\pm = \bphi_\pm \Delta t/2$, which is Eq. \eqref{eq:propagphipm}.

\section{Bounds of envelope drift \label{ap:fcdot}}
Here we study the variations of the function $g(.,.)$ defined in Eq. \eqref{eq:fcdot0}.
Using dummy variables $x,y$ as arguments, $g$ can be written as
\begin{equation}
\label{apeq:fcdotg}
        g(x,y) =- \frac{1+xy}{1+(x+y)^2}.
\end{equation}
We search for extrema with the condition $\partial_x g = \partial_y g =0$. Noting that the denominator of Eq. \eqref{apeq:fcdotg} is always strictly positive, straightforward manipulation leads to the system
\begin{eqnarray}
-y - 2x -x^2y+y^3 & = & 0, \\
-x - 2y -xy^2+x^3 & = & 0.
\end{eqnarray}
One trivial solution is given by $x=y=0$. It is in fact the only real solution and it is a global minimum as can be seen from the fact that $g(0,0) =-1$ and that the function asymptotically tends to 0 at infinity. 

In practice, in Eq. \eqref{eq:fcdot0} $y\in[-1,1]$. At any given $y \neq 0$ there are two extrema in $x$,
\begin{equation}
        x_{\pm} = \frac{-1 \pm \sqrt{y^4 - y^2 +1}}{y}.
\end{equation}
We may define the functions $g_{\pm}(x_{\pm}(y),y)$. These functions are necessarily monotonous functions of $y$ for $y \neq 0$ since we have seen that there exists only one real extremum of g at $x=y=0$. The function $g_{\pm}$ are even and they reach their maxima on the boundaries of the interval $[-1,1]$ with $g_{\pm}(1) = g_{\pm}(-1) = \mp 1/2$. The result follows that
\begin{equation}
        \forall (x,y) \in ]-\infty, +\infty[\times[-1,1], -1 \leq g(x,y) \leq \frac{1}{2}.
\end{equation}

\end{appendix}

\end{document}